%% file: main.tex
\documentclass[sigconf]{acmart}

\AtBeginDocument{%
  \providecommand\BibTeX{{%
    Bib\TeX}}}

\copyrightyear{2025}
\acmYear{2025}
\setcopyright{rightsretained}
\acmConference[ICPP '25]{54th International Conference on Parallel Processing}{September 08--11, 2025}{San Diego, CA, USA}
\acmBooktitle{54th International Conference on Parallel Processing (ICPP '25), September 08--11, 2025, San Diego, CA, USA}
\acmPrice{}
\acmDOI{10.1145/3754598.3754655}
\acmISBN{979-8-4007-2074-1/25/09}

\input{includes}
\input{defines}

\begin{document}

\input{titleauth}

\author{Yicong Luo}
\affiliation{%
  \institution{Georgia Institute of Technology}
  \city{Atlanta}
  \country{USA}
}
\email{yluo460@gatech.edu}

\author{Senhe Hao}
\affiliation{%
  \institution{Georgia Institute of Technology}
  \city{Atlanta}
  \country{USA}
}
\email{hsh@gatech.edu}

\author{Brian Wheatman}
\affiliation{%
  \institution{University of Chicago}
  \city{Chicago}
  \country{USA}
}
\email{bwheatman@gmail.com}

\author{Prashant Pandey}
\affiliation{%
  \institution{Northeastern University}
  \city{Boston}
  \country{USA}
}
\email{p.pandey@northeastern.edu}

\author{Helen Xu}
\affiliation{%
  \institution{Georgia Institute of Technology}
  \city{Atlanta}
  \country{USA}
}
\email{hxu615@gatech.edu}

\renewcommand{\shortauthors}{Yicong Luo, Senhe Hao, Brian Wheatman, Prashant Pandey, and Helen Xu}

\input{abstract}


\begin{CCSXML}
<ccs2012>
   <concept>
       <concept_id>10002951.10002952.10002971</concept_id>
       <concept_desc>Information systems~Data structures</concept_desc>
       <concept_significance>500</concept_significance>
       </concept>
   <concept>
       <concept_id>10010147.10010169.10010170.10010171</concept_id>
       <concept_desc>Computing methodologies~Shared memory algorithms</concept_desc>
       <concept_significance>500</concept_significance>
       </concept>
   <concept>
       <concept_id>10003752.10003809.10010031</concept_id>
       <concept_desc>Theory of computation~Data structures design and analysis</concept_desc>
       <concept_significance>500</concept_significance>
       </concept>
 </ccs2012>
\end{CCSXML}

\keywords{B-skiplist, blocked skiplist, concurrency}


\maketitle

\input{intro}
\input{prelim}
%\input{bskip-implementation}
\input{bskip-insert}
\input{bskip-concurrency}
\input{eval}
\input{conclusion}

\section*{Acknowledgments}
This research is funded in part by
NSF grant OAC 2339521 and 2517201.

\clearpage

\appendix
\section{Pseudocode}
\input{latexfigs/insert_algo}
\input{latexfigs/locking_logic}

\section{Deadlock Freedom}

\para{Deadlock-freedom} 
We will show that the proposed concurrent B-skiplist is deadlock-free.
The most popular way to avoid deadlock is to ensure that locks are acquired
in a total order, thereby avoiding circular wait. In the proposed top-down
concurrency control protocol, there is a total ordering on the locks from
left-to-right within a level, and then from top-to-bottom in the levels. That
is, the lowest ordered node is the left $-\infty$ sentinel at the highest level,
and the highest ordered node is the right $\infty$ sentinel at the leaf
level. The left $-\infty$ sentinel at some level $\ell$ is ordered directly
after the right $\infty$ sentinel at level $\ell + 1$. The insert and find
algorithms acquire locks according to this total ordering in a left-to-right and
top-to-bottom fashion, thereby avoiding deadlock.

\section{Zipfian Key Figures}

\input{latexfigs/zipfian_tp}

\input{latexfigs/zipfian_tp_tree}

\input{latexfigs/zipfian_latency}

\section{Data Tables}

\input{latexfigs/micros-table}

\input{latexfigs/ycsb_skiplist_table}

\clearpage

\balance

\input{latexfigs/ycsb_tree_table}

~\tabref{micros} contains the results of the parameter sweep that was used to determine the optimal parameter. The experiment was ran on 64 threads in a single NUMA socket.

~\tabref{skiplist-table} and ~\tabref{tree-table-small} contains the data used to generate figures~\ref{fig:skiplist-throughput}, ~\ref{fig:skiplist-latency}, ~\ref{fig:tree-throughput}, ~\ref{fig:tree-latency}, ~\ref{skipziptp}, ~\ref{treeziptp}, and ~\ref{ziplatency}.

\section{Artifact Instructions}

The \bskiplist implementation is available at \url{https://github.com/Ratbuyer/bskip_artifact}, including all the tested systems in Section 5. The README files in the repository contain compiling and running instructions.

\clearpage

\bibliographystyle{ACM-Reference-Format}

\end{document}

%% file: includes.tex
\usepackage{amsmath,amsfonts}
\usepackage{algorithm}
\usepackage{algpseudocode}
\usepackage{microtype}
\usepackage{graphicx}
\usepackage{textcomp}
\usepackage{xcolor}
\usepackage{xspace}
\usepackage{url}
\usepackage{amsthm}
\usepackage{balance}
\usepackage{tikz}
\usepackage{subcaption}
\usepackage{pgfplots}
\usepackage{multirow}
\usepackage{cleveref}

\pgfplotsset{compat=1.9}
\def\BibTeX{{\rm B\kern-.05em{\sc i\kern-.025em b}\kern-.08em
    T\kern-.1667em\lower.7ex\hbox{E}\kern-.125emX}}

\usepackage{enumitem}
\setitemize{leftmargin=*,noitemsep,topsep=0pt,parsep=0pt,partopsep=0pt}
\setenumerate{leftmargin=*,noitemsep,topsep=0pt,parsep=0pt,partopsep=0pt}

%% file: defines.tex
\definecolor{magenta4}{rgb}{0.5625,0,0.5625}
\definecolor{green4}{rgb}{0,0.5625,0}
\definecolor{orange4}{rgb}{0.98,0.31,0.09}
\newcommand{\prashant}[1]{{ \textcolor{green4}{Prashant: {#1}}}}
\newcommand{\eddy}[1]{{ \textcolor{orange4}{Eddy: {#1}}}}

\newcommand{\brian}[1]{{ \textcolor{purple}{Brian: {#1}}}}
\newcommand{\helen}[1]{{ \textcolor{magenta4}{Helen: {#1}}}}
\newcommand{\todo}[1]{{\textcolor{red}{TODO: {#1}}}}

\newcommand{\figref}[1]         {Figure~\ref{fig:#1}}

\newcommand{\tabref}[1]        {Table~\ref{tab:#1}}
\newcommand{\secref}[1]         {Section~\ref{sec:#1}}

\newcommand{\secreftwo}[2]      {Sections \ref{sec:#1} and~\ref{sec:#2}}
\newcommand{\figreftwo}[2]      {Figures \ref{fig:#1} and~\ref{fig:#2}}

\newcommand{\numthreads}{\ensuremath{128}\xspace}
\newcommand{\numcores}{\ensuremath{64}\xspace}
\newcommand{\nodesize}{\texttt{node\_size}\xspace}

\newcommand{\skiplist}{skiplist\xspace}
\newcommand{\skiplists}{skiplists\xspace}
\newcommand{\bskiplist}{B-\skiplist}
\newcommand{\bskiplists}{B-\skiplists}
\newcommand{\bskiplistshort}{BSL\xspace}

\newcommand{\btree}{B-tree\xspace}
\newcommand{\bplustree}{B+-tree\xspace}

\newcommand{\btrees}{B-trees\xspace}
\newcommand{\btreeshort}{OBT\xspace}
\newcommand{\Skiplist}{Skiplist\xspace}
\newcommand{\Skiplists}{Skiplists\xspace}
\newcommand{\schemename}{top-down\xspace}

\newcommand{\maxlen}{\texttt{max\_len}\xspace}
\newcommand{\defn}[1]{\textit{#1}}
\newcommand{\cslm}{ConcurrentSkipListMap\xspace}
\newcommand{\folly}{Folly\xspace}

\newcommand{\masstree}{Masstree\xspace}
\newcommand{\masstreeshort}{MT\xspace}

\newcommand{\javashort}{JSL\xspace}

\newcommand{\nohotshort}{NHS\xspace}
\newcommand{\follyshort}{FLY\xspace}
\renewcommand{\defn}[1]       {{\textit{\textbf{\boldmath #1}}}}

\newcommand{\bsloverjavathroughput}{$2 \times$}
\newcommand{\bslovernhsthroughput}{$11 \times$}
\newcommand{\bsloverflythroughput}{$2.1 \times$}
\newcommand{\bsloverjavatprange}{4.6-6.6$\times$\xspace}
\newcommand{\bslovernhstprange}{5-9$\times$\xspace}
\newcommand{\bsloverflytprange}{3.1-3.3$\times$\xspace}
\newcommand{\flyminlatencyratio}{$3.5 \times$\xspace}
\newcommand{\flycommonlatencyratio}{$3.7 \times$\xspace}
\newcommand{\bskipoverbtreerange}{1-1.4$\times$\xspace}
\newcommand{\bskipovermasstreerange}{1-2.1$\times$\xspace}
\newcommand{\bskipwithinbtree}{$0.9 \times$\xspace}
\newcommand{\bskipoverbtreeinsertsrange}{1.3-2.1$\times$\xspace\xspace}
\newcommand{\btreeoverbskipwklde}{$1.4 \times$\xspace}

\newcommand{\btreenodesize}{1024\xspace}
\newcommand{\bskipnodesize}{2048\xspace}
\newcommand{\bskipnodesizeelts}{128\xspace}
\newcommand{\scalingfactor}{0.5\xspace}
\newcommand{\promotionprob}{64\xspace}
\newcommand{\bskipoverslthroughputrange}{$2 \times$--$9 \times$\xspace}
\newcommand{\bskipoversllatencyrange}{$3.5 \times$--$103 \times$\xspace}

\newcommand{\bskipovertreethroughputpoint}{$0.9 \times$--$1.7 \times$\xspace}

\newcommand{\bskipovertreethroughputrange}{$0.7 \times$--$7.5 \times$\xspace}
\newcommand{\bskipovertreelatencyrange}{$0.85 \times$--$64 \times$\xspace}

\newcommand{\para}[1]{\smallskip\noindent\textbf{#1.}}

\usetikzlibrary{calc,matrix, shapes.multipart,patterns}
\pgfplotsset{
    discard if/.style 2 args={
        x filter/.append code={
            \edef\tempa{\thisrow{#1}}
            \edef\tempb{#2}
            \ifx\tempa\tempb
                
            \fi
        }
    },
    discard if not/.style 2 args={
        x filter/.append code={
            \edef\tempa{\thisrow{#1}}
            \edef\tempb{#2}
            \ifx\tempa\tempb
            \else
                
            \fi
        }
    },
}

\pgfplotsset{compat=1.8,
    /pgfplots/ybar legend/.style={
    /pgfplots/legend image code/.code={%
       \draw[##1,/tikz/.cd,yshift=-0.25em]
        (0cm,0cm) rectangle (3pt,0.8em);},
   },
}
\definecolor{safe-black}{RGB}{0,0,0}
\definecolor{safe-olive}{RGB}{0,73,73}
\definecolor{safe-teal}{RGB}{0,146,146}
\definecolor{safe-pink}{RGB}{255,109,182}
\definecolor{safe-peach}{RGB}{255,160,110}
\definecolor{safe-plum}{RGB}{73,0,146}
\definecolor{safe-cerulean}{RGB}{0,109,219}
\definecolor{safe-lavender}{RGB}{182,109,255}
\definecolor{safe-sky}{RGB}{109,182,255}
\definecolor{safe-baby}{RGB}{182,219,255}
\definecolor{safe-brick}{RGB}{146,0,0}
\definecolor{safe-brown}{RGB}{146,73,0}
\definecolor{safe-orange}{RGB}{219,209,0}
\definecolor{safe-green}{RGB}{36,255,36}
\definecolor{safe-yellow}{RGB}{255,255,109}

\pgfplotscreateplotcyclelist{mylist}{%
    {safe-pink, mark=square*},
    {safe-cerulean, mark=triangle*},
    {safe-peach, mark=x,mark options={scale=1.6}},
    {blue, mark=star,mark options={scale=1.6}},
    }

    \pgfplotscreateplotcyclelist{mylistshapes}{%
    {safe-pink, mark=square*},
    {safe-cerulean, mark=triangle*},
    {safe-peach,mark=*},
    {blue, mark=diamond*},
    }

%% file: titleauth.tex
\title{Bridging Cache-Friendliness and Concurrency:\\ A Locality-Optimized In-Memory B-Skiplist}

%%
%% The "author" command and its associated commands are used to define the
%% authors and their affiliations.
\iffalse
\author{Yicong Luo}

\affiliation{%
  \institution{Georgia Institute of Technology}
}
\email{yluo460@gatech.edu}

\author{Senhe Hao}
\affiliation{%
  \institution{Georgia Institute of Technology}
}
\email{hsh@gatech.edu}

\author{Brian Wheatman}
\affiliation{%
  \institution{University of Chicago}
}
\email{wheatman@uchicago.edu}
\author{Prashant Pandey}
\affiliation{%
  \institution{Northeastern University}
}
\email{p.pandey@northeastern.edu}

\author{Helen Xu}
\affiliation{%
  \institution{Georgia Institute of Technology}
}
\email{hxu615@gatech.edu}
\fi

%%% Local Variables:
%%% mode: latex
%%% TeX-master: "main"
%%% End:

%% file: abstract.tex
\begin{abstract}

% Both high cache locality and simple concurrency are essential for maximizing the performance of in-memory indexes, as they reduce cache misses and simplify parallel access, ensuring efficient data retrieval and updates.
% \skiplists offer easy implementation and simple concurrency control but lack cache locality, while \btrees offer high cache locality but come with complex concurrency control mechanism.

  \Skiplists are widely used for in-memory indexing
  % (i.e., the memtable)
  in many key-value stores, such as RocksDB and LevelDB,
  due to their ease of implementation and simple concurrency control mechanisms.
  % compared to \btrees.
  %Specifically, \skiplists offer robust worst-case latency guarantees for high-contention (i.e., insert-heavy) workloads.
  However, traditional \skiplists suffer from poor cache locality, as they store only a single element per node, leaving performance on the table.
  Minimizing last-level cache misses is key to maximizing in-memory index performance, making high cache locality essential.

  In this paper, we present a practical concurrent \bskiplist that enhances
  cache locality and performance while preserving the simplicity of
  traditional \skiplist structures and concurrency control schemes. Our key contributions
  include a top-down, single-pass insertion algorithm for \bskiplists and a corresponding
  simple and efficient top-down concurrency control scheme.

  % In this work, we introduce a practical concurrent
  % % \emph{simple}
  % \bskiplist that improves the cache locality and performance of \skiplists without trading off the simplicity of traditional \skiplist concurrency control scheme. The main novel algorithmic ingredients in this work are a top-down single-pass insertion algorithm for \bskiplists and a concurrency control scheme.

  On \numthreads threads, the proposed concurrent \bskiplist achieves between
  \bskipoverslthroughputrange higher
  throughput compared to state-of-the-art concurrent \skiplist implementations,
  including Facebook's concurrent skiplist from \folly and the Java
  \cslm. Furthermore, we find that the \bskiplist achieves competitive
  (\bskipovertreethroughputpoint) throughput on point workloads compared to
  state-of-the-art cache-optimized tree-based indices (e.g., Masstree). For a
  more complete picture of the performance, we also measure the latency of
  \skiplist- and tree-based indices and find that the \bskiplist achieves
  between \bskipoversllatencyrange lower 99\% latency compared to other
  concurrent \skiplists and between \bskipovertreelatencyrange lower 99\%
  latency compared to tree-based indices on point workloads with inserts.
  \end{abstract}
%%% Local Variables:
%%% mode: latex
%%% TeX-master: "main"
%%% End:

%% file: intro.tex
\section{Introduction}\label{sec:intro}

\iffalse
% Highly concurrent in-memory data storage systems (i.e., memtables) are becoming increasingly important with the rise of large multicore machines. Choosing an optimized scalable data structure is important for both throughput and latency, which are necessary for downstream applications~\cite{lersch2020enabling, dean2013tail, decandia2007dynamo, gregg2014systems}. In this paper, we will consider both \emph{latency} and \emph{throughput} as first-class citizens when measuring performance.
% %\helen{this sentence doesn't go here, but not sure yet where}

% In high-contention workloads with many threads, the primary performance bottleneck is not the actual workload but the overhead of the \defn{concurrency control} (CC) scheme~\cite{yu2014staring, bang2022full}, or the method that the index uses to manage simultaneous data accesses by multiple threads. Therefore, efficient thread-level parallelism via efficient CC protocols is the key to effectively scaling database management systems (DBMSs)~\cite{bang2022full, yu2014staring, MaoKoMo12, WangPaLi18}.
\fi

The \skiplist~\cite{pugh1990skip} has become a widely used in-memory index (i.e., the memtable) in many popular databases, including HBase~\cite{hbase}, RocksDB~\cite{rocksdb}, and LevelDB~\cite{LevelDB}. Additionally, Java features a \skiplist as its primary concurrent set and map implementation~\cite{javaskiplist}.

\iffalse
\Skiplists~\cite{pugh1990skip} are randomized self-balancing data structures
that support fast search, insertion, and deletion in $O(\log n)$ time \emph{with
  high probability}\footnote{Given an input size $n$, an event occurs with high
  probability if it holds with probability $1 - 1/n^c$ for some constant $c$.}
by layering multiple linked lists. The bottommost level is a standard sorted
linked list, while higher levels serve as "express lanes," allowing searches to
skip over multiple elements at a time. Each element is randomly (using coin
tosses) assigned a height upfront, determining how many levels it appears
in. Operations begin at the highest level, making large jumps initially and
refining the search as they descend. This structure provides \emph{logarithmic
  performance} similar to balanced trees (e.g., B-trees or AVL trees) but with
simpler implementation and efficient concurrent variants.\helen{maybe this para
  should just move to prelim?}
\fi

The main reason for the choice of \skiplists~\cite{pugh1990skip}
over trees (e.g., the \btree~\cite{BayerMc72}) is because \skiplists enable simple
structural modification operations. %(SMOs).
%and 2) simple and elegant concurrency
%control (CC) schemes.
In contrast to tree-based indices, \skiplists do not require
complex rebalancing operations %or internal adjustment operations
because elements are
randomly (using coin tosses) assigned a height upfront.
%(i.e., how many linked
%lists it appears in).
As a result, \skiplists support simple and effective (both
lock-based and lock-free) concurrency control (CC) schemes~\cite{herlihy2006provably,
  pugh1990concurrent, herlihy2007simple, fomitchev2004lock,
  fraser2004practical}.

\begin{table}[t]
\centering
\resizebox{\columnwidth}{!}{
\begin{tabular}{@{}cccccc@{}}
 \hline
 % & \multicolumn{2}{c}{B-tree} & \multicolumn{2}{c}{B-skiplist} \\
 \textit{YCSB Workload} & \textit{Skiplist} & \textit{\btree} & \textbf{\textit{\bskiplist}} & \textit{SL/BSL} & \textit{BT/BSL}  \\
  &  &  & (This paper) &  &   \\
 \hline
 % took the numbers from ycsb sheet if they were there, else the uniform sheet
 Load + C & 4.9E9  & 2.1E9 & 1.5E9 & 3.2  & 1.4 \\
 Load + E & 1.1E10 & 2.3E9 & 2.0E9 & 5.6  & 1.2 \\
 \hline
\end{tabular}
}
\caption{LLC load misses\protect\footnotemark~of Facebook's folly \skiplist (SL)~\cite{folly}, a concurrent
\btree (BT)~\cite{XuLiWh23} and the concurrent \bskiplist (this paper, BSL) during the YCSB~\cite{ycsb}
  load and run phase. The load phase has 100\% inserts, workload C has 100\%
  finds, and E has 95\% range queries/5\% inserts.} %\todo{update with 128 threads and change ratios downstream}}
%The last two columns show the ratio of the cache misses for \skiplist and \btree over \bskiplist.}
\label{tab:intro-cache}
\end{table}

\para{Locality issues in \skiplists} Unfortunately, traditional \skiplists exhibit poor spatial locality because they store a single element per node. In
contrast, cache-friendly indexes such as B-trees~\cite{BayerMc72} store multiple
elements per node, reducing the height of the structure and therefore the number of memory fetches during top-to-bottom traversals. Furthermore, storing
multiple elements per node further reduces cache misses during horizontal traversals.
%during range scans.

~\tabref{intro-cache} shows that \skiplists incur significantly more cache
misses than \btrees, leaving performance on the table. Concretely, on the tested
workloads, a state-of-the-art skiplist~\cite{folly} incurs $2.4-4.8\times$ more
cache misses than a comparable \btree~\cite{XuLiWh23} on both point and range
workloads. \footnotetext{\secref{eval} contains all details about the
  experimental setup.}  This discrepancy in cache misses translates into actual
performance: the \btree achieves between $2\times$-$8\times$ higher throughput
than the \skiplist on these workloads.

% \prashant{The above paragraph is a bit convoluted. Need to say that LLC misses dominate performance. Though simple and ameanable to concurrency Skiplists are leaving performance on the table.}

%Nevertheless, the
%\skiplist remains a popular choice for the in-memory index component of
%key-value stores due to its simplicity both in terms of structure and CC
%schemes.
\emph{In this paper, we introduce a \defn{concurrent \bskiplist} that improves the
cache locality (and therefore the performance) of the traditional \skiplist without giving up on the \skiplist's simple and effective CC schemes. Furthermore, due to its simplified CC scheme, the \bskiplist achieves lower worst-case latency than \btrees.}

% \brian{could this be left for the preliim section?}
\para{\Skiplist structure}
% To understand the proposed \bskiplist, we must first overview the structure of a traditional (non-blocked) \skiplist.
At a high level, \skiplists~\cite{pugh1990skip} are randomized self-balancing
data structures that support fast search, insertion, and deletion in $O(\log n)$
time \emph{with high probability}\footnote{An event $E_n$ on a problem of size
  $n$ occurs \defn{with high probabilty} if $Pr[E_n] \geq 1 - 1/n^c$ for some
  constant $c$.} (w.h.p.)  by layering multiple linked lists. The bottommost level is a
standard sorted linked list, while higher levels serve as ``express lanes,''
allowing searches to skip over multiple elements at a time.

The \skiplist is parameterized by a \defn{promotion probability} $p$, which
determines the likelihood of elements appearing in higher levels of linked
lists.  Upon insertion, an element is randomly (using coin tosses with
probability of heads $p$) assigned a height equal to the number of successive
coin tosses until heads for that particular element. The height of an element is
the number of linked-list levels that element appears in, starting from the
lowest level.

\para{Theoretical steps towards a cache-optimized \skiplist} We will use the I/O
model (or external-memory model)~\cite{AggarwalVi88}, which measures how well an
algorithm or data structure takes advantage of spatial locality by measuring
cache-line transfers. The model is parameterized by a cache-line size
$Z$. Transferring $Z$ contiguous elements in one cache line has unit cost in the
model.

Theoretically, given some node size $B=\Theta(Z)$, the straightforward way to
improve the cache-friendliness of \skiplists is to promote elements with
probability $1/B$ rather than $1/2$ as in traditional \skiplists.  Indeed,
Golovin~\cite{golovin2010b} proposed this method with the \defn{\bskiplist}, a
cache-optimized \skiplist that matches the \btree's bounds \emph{in
  expectation}. For each level in this theoretical \bskiplist, consecutive
unpromoted elements are stored in the same node. However, this initial paper on
\bskiplists stops short of providing  theoretical guarantees w.h.p.,
 parallelization, and implementation, leaving a massive gap in making
\bskiplists practical.

% \brian{I feel like these next two sections could be tightened up}
\para{Challenges to blocking \skiplists} In practice, addressing
locality issues in \skiplists via blocking (i.e., storing multiple elements per
node) raises challenges due to the randomized \emph{variable size of
  nodes}. Theoretically, each node in a \bskiplist has $\Theta(B)$ elements in
expectation, but there exist nodes with as many as $\Theta(B \log n)$ elements
w.h.p.~\cite{bender2016anti, bender2017write}. Since nodes can be large, finds
and inserts, which require scanning and potentially shifting a linear number of
elements in a node, can cost as much as $O(\log n)$ cache-line transfers in the
I/O model, matching a regular skiplist's randomized bounds.

% The naive solution of table doubling (either in-place or out-of-place)
% to store all elements in a node in one contiguous block of memory would require both $O(\log n)$ cache-line reads and writes due to element shifts during insertions w.h.p. Since there exist nodes with $\Theta(B \log n)$ elements, it would take $\Theta(\log n)$ cache misses to insert an element in such a node (given $B = \Theta(Z)$).
% \prashant{We can probably comment the above paragraph. It is not adding much extra info.}

\para{Bounding the element moves in a \bskiplist} To mitigate this issue in practice,
we enforce \defn{fixed-size} physical nodes in the \bskiplist to bound the
maximum number of element moves during insertions. \bskiplist nodes are allowed
to grow to arbitrary sizes according to the results of randomized
promotions. However, if a logical \bskiplist node contains $k > B$ elements, we
physically store it as $\lceil{k/B}\rceil$ nodes at the same level connected by
pointers.

The design choice of fixed-size nodes is subtle but is key to fast inserts in
practice because it limits the worst-case number of cache-line writes to
$\Theta(1)$ per level. In contrast, if physical nodes are allowed to grow
arbitrarily, the maximum number of cache-line writes in a B-skiplist node is $O(\log n)$
w.h.p. To be clear, this choice does not affect the overall insertion bound in
the I/O model as each insert still requires $O(\log n)$ cache-line reads
w.h.p. to determine the correct position. However, as we shall see in the
empirical evaluation, fixed-size nodes enable the \bskiplist to support fast
insertions and low variance in the latency of operations.

%\brian{I think this should be renamed to make it more clear this is the comparison to exsistinng work and how they don't quite do what they claim to do}
\para{Prior steps towards concurrent blocked \skiplists} Furthermore, any
candidate for the in-memory index in databases, including the proposed
\bskiplist, must be \emph{concurrent} to take full advantage of parallel
resources in today's multicore machines.

On the practical side, there have been several steps towards improving locality in concurrent \skiplists\cite{xie2017parallelizing, na2023esl}, but these often
involve periodic rebuilding of the upper levels, % (either synchronously or
% asynchronously),
increasing the worst-case latency of individual operations. For example,
CSSL~\cite{SpreZeLe16} and PI~\cite{xie2017parallelizing}
 periodically rebuild the upper levels of the index, blocking reads and writes during the restructuring process.

 Other blocked \skiplists vary the component data structures and CC schemes at
 different levels, giving up on the simplicity of the original concurrent
 \skiplists.  For example, ESL~\cite{na2023esl} first inserts elements only into
 the bottom level and later updates the upper index levels asynchronously with
 background threads. The ESL is composed of two levels with distinct index
 structures and CC mechanisms.
%letting inserts to only immediately modify bottom layers and uses background threads and an operation log to update the upper index levels asynchronously.
Similarly, S3~\cite{ZhangWuTa19}, another cache-sensitive \skiplist, employs a similar strategy of a two-level index with different CC schemes. Furthermore, it adaptively chooses ``guard entries'' with a neural model, giving up on the randomization  of \skiplists and therefore the \skiplist's probabilistic theoretical guarantees.

\para{Designing a concurrent \bskiplist} In this work, we focus on designing a
\emph{simple} and \emph{effective} CC mechanism for \bskiplists without
modifying its high-level structure. Specifically, we propose a CC mechanism
based on fine-grained locking to achieve both high throughput and low worst-case
latency. The ideal CC scheme for \bskiplists inherits the simplicity and
performance of \skiplist-based CC schemes~\cite{pugh1990concurrent,
  herlihy2006provably, fomitchev2004lock, herlihy2007simple}, so we use them
as a starting point. However, they do not %immediately extend to \bskiplists as
% they rely heavily on pointer manipulation and
handle node splits and merges.

%\helen{this transition needs smoothing}
At a high level, \skiplist insertion algorithms (and their corresponding CC mechanisms) make two traversals through the index if an element is promoted: one ``top-down'' read-only phase to
determine the location(s) that an element should be inserted, and then a
corresponding ``bottom-up'' write-only phase that links in the new \skiplist
nodes~\cite{pugh1990concurrent, herlihy2006provably, fomitchev2004lock, herlihy2007simple, platz2019concurrent}. In both the top-down and bottom-up phase, a CC scheme for \skiplists may either 1) hold a constant number of locks via hand-over-hand locking~\cite{pugh1990concurrent} or 2) hold locks on all levels that an element is promoted to~\cite{herlihy2006provably, fomitchev2004lock, herlihy2007simple, platz2019concurrent}.
%\prashant{We should validate this claim.}
%These schemes have been shown to achieve good performance and scale well.

To reduce the number of traversals in both sequential and concurrent insertions, we introduce a \defn{top-down} insertion\footnote{We focus on the case of insertions due to space constraints, but deletions are symmetric.} algorithm and %corresponding CC scheme for \bskiplists
that exploits inherent \skiplist
properties to traverse the \bskiplist
\emph{only once} (top-to-bottom and left-to-right). In contrast, existing \skiplist insertion algorithms make two traversals if an element is promoted.
As we shall see, the design of the insertion algorithm
% is subtle and
has
major implications for the corresponding CC mechanism.
% The insertion algorithm and CC scheme are tightly coupled in any data structure, not just \bskiplists, because the insertion algorithm determines the order and number of locks that must be acquired in the corresponding CC scheme.

We build upon this top-down insertion algorithm to develop a \emph{simple} yet \emph{efficient} \schemename
CC scheme based on reader-writer
locks~\cite{courtois1971concurrent}, a common synchronization primitive for in-memory indexes. The \schemename CC scheme minimizes overheads by
1)
avoiding multiple retires from root-to-leaf, and 2) minimizing both the number
of exclusive locks held at one time and the duration that the locks are held.
Using the \schemename CC scheme, a thread only needs to lock nodes on at most two layers at once, and locks only a constant number of nodes at once.

\para{Contributions} The contributions of the paper are as follows:
\begin{itemize}[leftmargin=*,noitemsep]
\item The design of the \bskiplist with fixed-size nodes to improve cache locality and mitigate worst-case behavior.
\item A novel top-down concurrency control mechanism for \bskiplists built on a corresponding insertion algorithm for \bskiplists that completes
  insertions in one pass from top-to-bottom.
\item A C++ implementation of the concurrent \bskiplist.
\item An empirical evaluation of the concurrent \bskiplist compared to several
  in-memory skiplist-based and B-tree-based indexes that demonstrates that the \bskiplist achieves between \bskipoverslthroughputrange higher throughput on YCSB workloads~\cite{ycsb} compared to state-of-the-art concurrent \skiplists.
\end{itemize}

%\brian{maybe I have just been thinking about this problem for too long, but is top down that novel, I didn't think we came up with it?}
%\prashant{Top down is not novel. We should not claim that. This is same as latch crabbing or hand over hand locking without restarts.}\helen{top down is novel in the context of skiplists. maybe we can clarify}

\input{latexfigs/ycsb_intro_plot}

\para{Results summary}~\figref{skiplist-throughput} demonstrates that the
\bskiplist achieves between \bskipoverslthroughputrange higher throughput on
workloads from the popular Yahoo! Cloud Serving Benchmark (YCSB)~\cite{ycsb} compared to state-of-the-art
concurrent \skiplists including Facebook's concurrent \skiplist from the \folly
library~\cite{folly}, the Java \cslm~\cite{javaskiplist}, and the No Hot Spot
Skiplist (NHS)~\cite{crain2013no}.
% Notably, the \bskiplist achieves $3.1\times$-$6.1\times$ speedup over the
% optimized concurrent \skiplist from the popular \folly library.
  The YCSB workloads include a mix of point operations (finds/inserts) and range operations.
  % Due to the \bskiplist's cache-friendliness as shown in~\tabref{intro-cache},
  % it achieves higher throughput on all workloads compared to non-blocked
  % concurrent \skiplists.

Furthermore, we also evaluate two tree-based indices: a state-of-the-art concurrent \btree~\cite{XuLiWh23} and Masstree~\cite{MaoKoMo12}, a popular cache-optimized in-memory index. Compared to tree-based indices, we find that the \bskiplist achieves competitive throughput (between \bskipovertreethroughputpoint) on point workloads, and between \bskipovertreethroughputrange throughput on range workloads.

In addition to throughput, we evaluate all data structures on latency as well
for a more complete picture of data-structure performance. As we shall detail
in~\secref{eval}, the \bskiplist achieves between \bskipoversllatencyrange lower
99\% latency compared to existing concurrent \skiplists.
%A \btree~\cite{XuLiWh23} can achieve up to $1.2\times$ lower 99th percentile latency on read-heavy workloads compared to the \bskiplist. On the other hand,
%the \bskiplist supports between \bskipovertreelatencyrange lower 99th percentile latency on point workloads with inserts compared to state-of-the-art tree-based indices.

%%% Local Variables:
%%% mode: latex
%%% TeX-master: "main"
%%% End:

%% file: latexfigs/ycsb_intro_plot.tex
\begin{figure}
  \centering
  \begin{tikzpicture}
    \begin{axis}[
    width=8cm, height=3.5cm,
        ybar,
        legend style={at={(0.5, 1.4)}, anchor=north},
        ylabel={Normalized throughput},
         ylabel style={align=center, text width=3cm},
        xlabel={Workload},
        xtick=data,
        bar width=5pt,
        legend columns=5,
        ytick={1,5,10},
        enlarge x limits=0.14,
        symbolic x coords={Load, A, B, C, E},
        x tick label style={text width = 1cm, align = center},
        ymin = 0,
        ymax = 12,
    %    extra y tick style={
    %                 % in case you should remove the grid from the "normal" ticks ...
    %                 ymajorgrids=true,
    %                 % ... but don't show an extra tick (line)
    %                 ytick style={/pgfplots/major tick length=0pt,},
    %                 grid style={violet ,dashed,},
    %         },
    % every axis plot/.append style={fill},
      ]

      \addplot coordinates {
    (Load,1)(A, 1)(B, 1)(C, 1)(E, 1)
    };
      \addlegendentry{NHS}

      \addplot table [
          x=workload, y=folly_ratio, col sep=tab
        ] {tsvs/ycsb_uniform_tp.tsv};
      \addlegendentry{Folly SL}
    \addplot [fill=safe-lavender, postaction={pattern=north east lines}] table [
          x=workload, y=javaskip_ratio, col sep=tab
        ] {tsvs/ycsb_uniform_tp.tsv};
        \addlegendentry{Java SL}
            \addplot [fill=safe-teal, postaction={pattern=crosshatch}] table [
          x=workload, y=NHS_ratio, col sep=tab
          ] {tsvs/ycsb_uniform_tp.tsv};
          \addlegendentry{\bskiplist}
    \end{axis}
  \end{tikzpicture}
  \vspace{-.5cm}
  \caption{Normalized throughput (ops/s) of \skiplist-based indices relative to
    the No Hot Spot Skip List(NHS) ~\cite{crain2013no}.}
    %A number below 1 means that the index achieved lower throughput than the \bskiplist. \brian{remove second sentance}
    \label{fig:skiplist-throughput}
  \end{figure}
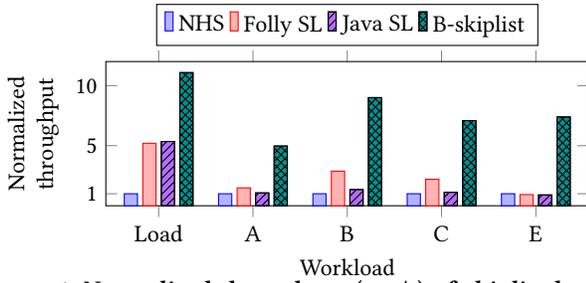

%%% Local Variables:
%%% mode: latex
%%% TeX-master: "../main"
%%% End:

%% file: prelim.tex
\section{Preliminaries}\label{sec:prelim}

This section will give background on the structure and operations of \skiplists
and \bskiplists necessary to understand later sections. For the operations, it
will focus on inserts for simplicity, but deletes are symmetric. Furthermore, it
will review the reader-writer synchronization primitive, which is the core
functionality that the proposed top-down single-pass concurrency control scheme
for \bskiplists is based on.

% Due to space limitations, we will not go into the details of \btree-based
% concurrency schemes, but we refer the interested reader to a survey on \btree
% locking techniques~\cite{graefe2010survey}.

\para{Operations}
A key-value dictionary data type stores pairs of keys and values \texttt{(k,
  v)}. Their main operations are as follows:
\begin{itemize}
\item \texttt{find(k)}: return the associated value \texttt{v}.
\item \texttt{insert(k, v)}: add \texttt{(k, v)} to the data
  structure.
\item \texttt{range(k, f, length)}: apply the function \texttt{f} to the
  \texttt{length} key-value pairs with the smallest keys that are at least
  \texttt{k}.
\end{itemize}
We consider these operations since they comprise the popular YCSB~\cite{ycsb}
workloads we use to perform the evaluation in~\secref{eval}.

\subsection{Skiplist structure and operations}
\begin{figure}[t]
  \centering \includegraphics[width=.8\linewidth]{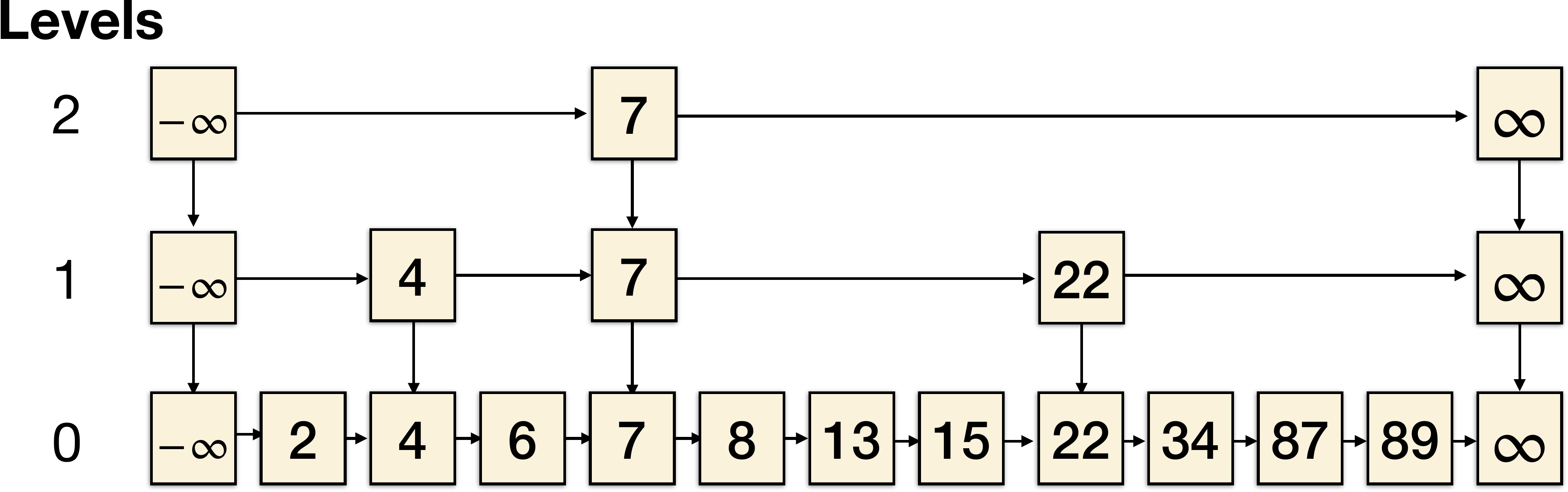}
  \vspace{-.3cm}
  \caption{Example of a \skiplist.}
  \label{fig:sl-example}
\end{figure}
\para{Structure} The \skiplist~\cite{pugh1990skip} is a data structure that
stores a \emph{hierarchy} of levels of linked lists. Each linked list is sorted
by the keys of the elements in that linked list, with sentinels at the beginning
and end for $-\infty$ and $\infty$. The bottommost level (level 0) contains all
of the elements in the data structure. The list size decreases by a constant
factor in expectation at each successive level as we move up the hierarchy. Each
linked list at some level $\ell > 0$ contains a subset of the elements in the
linked list below it (level $\ell - 1$). This property is called \emph{inclusion
  invariant}, where every element present at level $\ell$ must also be present
at levels $0, \ldots, \ell - 1$.

\figref{sl-example} illustrates the skiplist's pointer structure.  Just like in
trees, we will refer to the nodes at the lowest level of the \skiplist
($\ell = 0$) as \emph{leaf} nodes and those at higher levels ($\ell > 0$) as
\emph{internal} nodes. All nodes have a \texttt{next} pointer to a successor
node in the same level, and all internal nodes also have a \texttt{down} pointer
to the node with the same key in the level below. The \texttt{root} node is the
leftmost node at the highest level.

An element that appears at some maximum level $\ell > 0$ is said to be
\defn{promoted} to that level. Promotions are determined upfront at the start of
an insertion via randomization with a series of coin flips. Notably, the level
that an element is promoted to in a \skiplist is unrelated to the current
structure of elements in the \skiplist. That is, the highest level that an
element appears at in a \skiplist is equal to the number of successive ``heads''
seen when flipping a coin with some constant probabilty $p$ (usually $p = 1/2$,
but any probabilty $1/c$ where $c$ is a constant sufficies to achieve the
asymptotic bounds). Given a \skiplist with $n$ elements, the maximum height of
any element is $O(\log n)$ in expectation and with high probability.

\para{Operations}
All operations in a \skiplist start at the upper left sentinel ($-\infty$) at the
highest level and traverse through the pointer structure in a left-to-right and
top-to-bottom fashion. The skiplist is a self-balancing data structure and does
not require pointer rotation to maintain its bounds. A skiplist with $n$
elements supports all point operations (insert, delete, find) in $O(\log n)$
time in expectation and with high probability.

For ease of understanding, given a node with key \texttt{k} at a given level
$\ell$, let its \texttt{pred} (predecessor) element be the largest key strictly less than
\texttt{k}, and let its \texttt{succ} (successor) element be the smallest key strictly
greater than \texttt{k} at level $\ell$.  Except for the beginning and end
sentinels, all nodes in a \skiplist have logical \texttt{pred} and \texttt{succ}
elements which correspond to nodes in the skiplist.

To \texttt{find} a key \texttt{k}, the traversal searches left-to-right
starting from the left sentinel on the highest level. Upon finding its
\texttt{succ} node at that level, the search follows the down pointer of the
\texttt{pred} node. The search continues in this way until the lowest level,
where it scans left to right until it either finds \texttt{k} or does not find
it and encounters some element greater than \texttt{k}.

The \texttt{range} operation is a direct extension of a \texttt{find}. Rather
than terminating at the search key \texttt{k} or its \texttt{succ} element in
the leaf level, the \texttt{range} search continues a left-to-right search
through the leaf layer until \texttt{length} elements have been read, or the
search reaches the rightmost sentinel.

Inserts are similar to finds in terms of traversal order, but must link in a new node
containing some key \texttt{k} at each level that it is promoted to. Let
$h \geq 0$ be the height that the newly inserted element is promoted to. One way
to perform this structural modification is to keep track of the \texttt{pred}
nodes of \texttt{k} at all levels during the downward traversal. After the
search reaches the bottom level, the insert operation creates $h + 1$ new nodes
containing \texttt{k} and adjusts the \texttt{next} pointers at each level so
that \texttt{k->next = pred->next} and then \texttt{pred->next =
  k}. Furthermore, the insert must update the down pointers in each internal
node with the key \texttt{k} to the node containing \texttt{k} in the next
lowest level.

\subsection{\bskiplist structure and operations}
\begin{figure}[t]
  \centering \includegraphics[width=\linewidth]{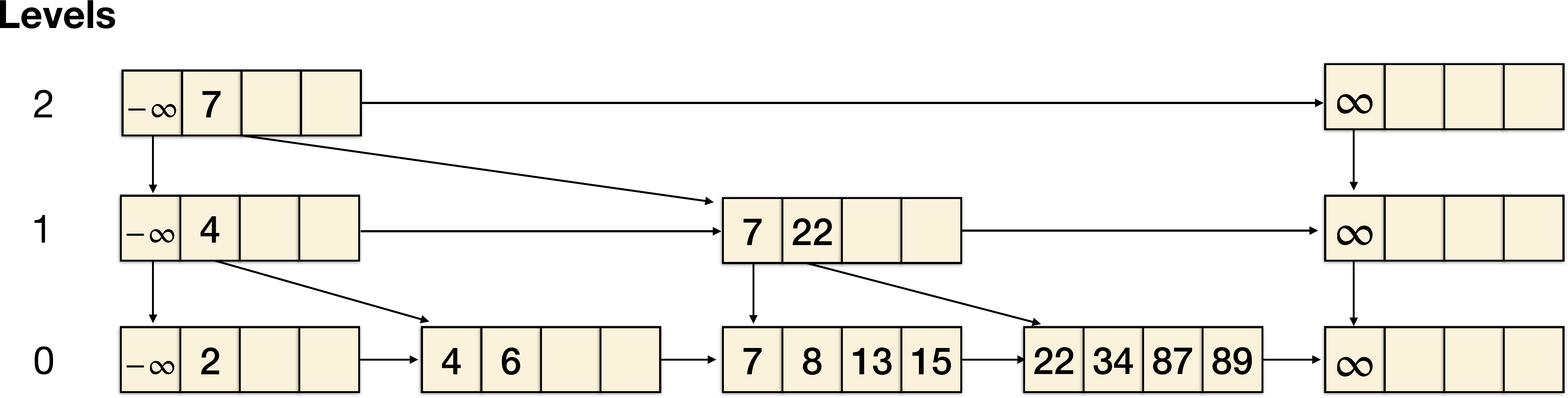}
  \vspace{-.75cm}
  \caption{Example of a \bskiplist with node size $B = 4$.}
  \label{fig:bskip-example}
\end{figure}

\para{Structure} The \bskiplist is a ``blocked'' version of a \skiplist that
contains multiple elements per node~\cite{golovin2010b}. The elements are
totally sorted at each level of the B-skiplist (by pointer structure and within
nodes). Given a cache-line size $Z$ and desired node size $B = \Theta(Z)$, the
\bskiplist's promotion probability is generally set to $p = \Theta(1/B)$.  In
theory, the expected number of elements per node is $\Theta(B)$, but in a
\bskiplist with $n$ elements, the maximum number of elements in a node is
$\Theta(B\log n)$ w.h.p.~\cite{bender2016anti, bender2017write}. Golovin's
original theoretical paper on \bskiplists~\cite{golovin2010b} does not address
how to handle this case in practice, but one straightforward solution is
table doubling when the node becomes full.

The block structure depends on promoted height of each element in the
B-skiplist. The \defn{header} key of a node in a \bskiplist is the first (and
smallest) element in that node. By construction, every header in a node at some
level $\ell$ has been promoted to level $\ell + 1$. %because it causes a split.

~\figref{bskip-example} illustrates the node and pointer structure of a
\bskiplist. Just as in a \skiplist, each node has a \texttt{next} pointer
pointing to the next node. Given a node \texttt{x}, all keys in \texttt{x} are
smaller than the header key of \texttt{x->next}. Furthermore, each internal node
(at level $\ell > 0$) contains an array of $B$ \texttt{down} pointers (one per
key in the node), each pointing to the corresponding node at the level below.

\para{Bounds} In the external-memory model described in~\secref{intro}, given a
cache-line size $Z$, a \bskiplist with $n$ elements and node size
$B = \Theta(Z)$ supports finds and inserts in $\Theta(\log_Z(n))$ cache-line
transfers in expectation (matching B-tree bounds). Furthermore, range queries
with $r$ elements in the range take $\Theta(\log_Z(n) + r/Z)$ cache-line
transfers in expectation. However, finds and inserts take $\Theta(\log n)$
cache-line transfers in the worst case w.h.p., matching the bounds of a
non-blocked \skiplist.

\para{Operations} The
% high-level
left-to-right and top-to-bottom pointer
traversal in a \bskiplist during finds and inserts is similar to the traversal in
a \skiplist. The main difference is that traversals in a \bskiplist must look at multiple
elements within a single node and determine which \texttt{down} pointer to
follow at an internal nodes.

To \texttt{find} a key \texttt{k} in a \bskiplist, the traversal begins at
a \texttt{curr} node initialized to the upper left sentinel (with header
$-\infty$), just as in the regular skiplist. The search then examines the header
of \texttt{curr->next}. If it is less than \texttt{k}, the \texttt{curr} node is
updated to \texttt{curr->next}. We repeat this left-to-right traversal until the
header of \texttt{curr->next} is greater than \texttt{k}. At that point, we
search within \texttt{curr} for the \texttt{pred} element and follow its
\texttt{down} pointer. The search continues in this way until we reach the leaf
level, at which point we determine if \texttt{k} is present.  The main
difference from a traditional \skiplist is that the \texttt{pred} element may be
in the same node as \texttt{k}.

To \texttt{insert} an element in a \bskiplist, we first perform a
\texttt{find}-like traversal to find all the \texttt{pred} elements at each
level. However, the node-modification operations during a \bskiplist insert are
similar to those in a B-tree. Let $h$ denote the height at which the key to be
inserted \texttt{k} is promoted to. If it not promoted (i.e., $h = 0$), we can
simply add it to the same node its \texttt{pred} element resides in and shift
all subsequent elements in the node one slot down. However, if it is promoted
(i.e., $h > 0$), we must perform a \texttt{split} at levels
$\ell = 0, 1, \ldots, h - 1$. Let \texttt{old\_node} be the node with
\texttt{pred}. A node \texttt{split} in a \bskiplist creates a
\texttt{new\_node} with \texttt{k} as the header and copies all elements (and
their down pointers, if the split occurs at an internal node) in
\texttt{old\_node} greater than \texttt{pred} (and also greater than \texttt{k})
after \texttt{k}. To link in the new node, we set \texttt{new\_node->next =
  old\_node->next} and \texttt{old\_node->next = new\_node}.

\para{Open problem} Golovin's paper on \bskiplists does not address the
\emph{order} of levels in which an element is inserted (i.e., starting from the
top or the bottom)~\cite{golovin2010b}. However, most traditional skiplist
insertion algorithms insert elements in a ``bottom-up'' fashion - they link in
new nodes starting from the leaf level up until the maximum level an element is
to be promoted to~\cite{pugh1990skip, pugh1990concurrent, herlihy2006provably,
  herlihy2007simple}. A similar algorithm for B-skiplists would perform a find
to the bottom level, keeping track of the affected nodes along the way, and
insert the element starting from the leaf level, performing splits and adjusting
pointers on the way up as necessary to always maintain the inclusion invariant.

\iffalse
Just as in traditional skiplists, exis ``Bottom-up'' algorithms first perform a
search from the top leftmost element down to the correct node in the leaf level
and keep track of the appropriate node that contains the predecessor element at
each level, as described in~\secref{prelim}. For concreteness, suppose we are
inserting an element \texttt{k} into a \bskiplist that will be promoted to some
height $h$ (that is, \texttt{k} will appear in all levels $\ell = 0, 1, \ldots,
h$).
\fi

\subsection{Concurrency control primitives}

\para{Hand-over-hand locking} Next, we will review the classical
\emph{hand-over-hand} (HOH) fine-grained locking scheme (also known as \emph{latch crabbing}
in databases) for sorted singly-linked lists~\cite{herlihy2020art},
%~\cite{herlihy2020art, hohmagazine,
%arpaci2018operating, scott2000programming}
which we will be building upon in
later sections.  HOH traverses a list
while holding at most two locks at a time, starting
%In HOH locking, traversals start by
at the \texttt{head} node of the linked list, then acquiring the lock on
the successor before releasing the lock on the predecessor. After locking a
node \texttt{curr}, it is safe to access \texttt{curr->next} to either perform a
comparison with a target key (e.g., for a search), or to see whether we have
reached the end of the list. During inserts, a traversal must hold two locks at
a time to link in a new node in the correct place between a node
\texttt{curr} and \texttt{curr->next}. HOH locking can also be extended to
\skiplists because \skiplists are simply towers of linked lists. 

\para{Reader-writer locks}
Finally, we will review reader-writer locks (RW
locks)~\cite{courtois1971concurrent}, as we will use them as the core
synchronization primitive in the proposed concurrent B-skiplist. RW locks enable
concurrent access for read-only operations but exclusive access for write
operations. Given a RW lock, a thread can either call \texttt{read\_lock()} to
access it in shared read mode, or \texttt{write\_lock()} to access it in
exclusive write mode. Multiple threads can read the data in parallel if they all
hold the lock in read mode, but all other threads must wait if a thread holds
the lock in write mode.

RW locks are the foundation of many concurrency control protocols in \btrees,
including the classical optimistic concurrency control (OCC)~\cite{KungRo81}.
%As a result, significant research has been devoted to optimizing RW-lock
%performance and scalability~\cite{mellor1991scalable, calciu2013numa,
%  lev2009scalable, liu2014scalable}.

%%% Local Variables:
%%% mode: latex
%%% TeX-master: "main"
%%% End:

%% file: bskip-insert.tex
\section{Top-Down Insertion Algorithm}\label{sec:bskip-insert}

This section presents the \defn{top-down} insertion algorithm that will form the
basis of the concurrent \bskiplist in the next section. We will first describe the algorithm serially and show that it
results in the same \bskiplist structure as the original ``bottom-up'' insertion
algorithm. In the next section, we will introduce a corresponding top-down
concurrency control scheme. Furthermore, we will show that the proposed
insertion algorithm achieves the same asymptotic runtime bounds of the
theoretical \bskiplist. For simplicity, we will describe the algorithm for keys only, but
 storing associated values involves only updating the leaf level.  We will
 initially describe the insertion algorithm using logical nodes for simplicity
 and then explain how to adapt it with fixed-size physical nodes.
%, but adapting it to
%fixed node sizes simply requires splitting a single physical node due to
%promotion.

\begin{figure}[t]
  \centering \includegraphics[width=\linewidth]{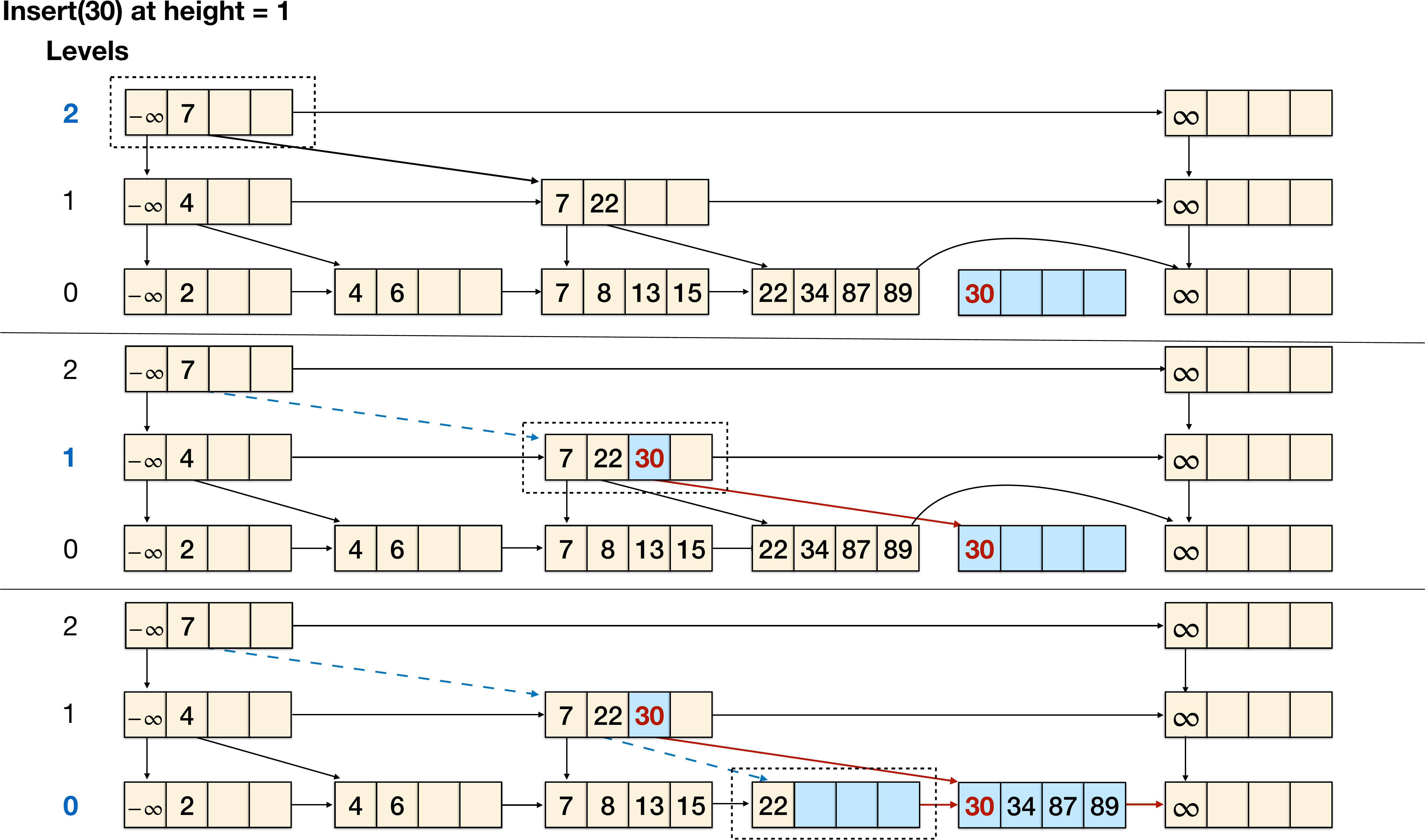}
  \caption{Example of a top-down insertion of key \texttt{k} = 30 and
    height $h = 1$ into a B-skiplist with node size $B = 4$. The dotted box
    represents the current node visited during the traversal, the blue dashed
    lines denote the pointers followed during the top-down traversal, and the
    blue cells represent the cells written during the insertion.}
  \label{fig:bskip-insert}
\end{figure}
\para{Description}
At a high level, the main change in the proposed ``top-down'' insertion
algorithm compared to existing ``bottom-up'' algorithms is the order in which
new elements are added to the skiplist. Although the distinction about the order
of insertion into levels is subtle, it has important implications for the
traversal direction, and as we shall see in~\secref{concurrency}, for CC mechanisms.

The goal of the top-down algorithm is to complete insertions in \emph{one pass}
of the skiplist without revisiting nodes by taking advantage of the skiplist
property that the height at which an element is promoted to is determined
\emph{upfront} and is \emph{independent of the current structure}. In contrast
to promotions in a B-tree, which depend on the current structure and fullness of
the nodes, element promotions in a B-skiplist depend only on a sequence of
random coin flips.

Rather than traversing down to the leaf level and then linking in elements from
bottom to top, as in most skiplist insertion algorithms, we propose to insert
elements starting from the highest level to which they are promoted and ending
with the leaf level. Therefore, an insertion is finished once the traversal
reaches the leaf level and performs a write at the relevant node.

% In contrast, as described in~\secref{prelim}, a ``bottom-up'' insertion
% algorithm for B-skiplists would first insert the element at the leaf level and
% work its way up the levels, performing splits as necessary until the level that
% the element should be promoted to. Such an algorithm 
% requires a ``top-down''
% read-only pass to the leaf level, and then a ``bottom-up'' write pass from the
% leaves upwards to insert the element into any relevant level(s). With a
% bottom-up strategy, an insertion is not necessarily complete once the traversal
% reaches the leaf level, since the element may
% need to
% be promoted to higher levels.

Suppose we are inserting an element \texttt{k} that will be promoted
to level $h \geq 0$ as determined by random coin flips. We traverse the
B-skiplist from top to bottom starting at the leftmost sentinel at the top level
as described in~\secref{prelim}. If the traversal is currently on some level
$\ell > h$, the element \texttt{k} will not appear on $\ell$, so the search
order is exactly the same as an original \bskiplist \texttt{find}. By following
\texttt{next} and \texttt{down} pointers, we will eventually reach level $h$,
where we will write \texttt{k} into the same node as its predecessor
\texttt{pred} at level $h$. If $h = 0$, the insertion is complete.

The main challenge comes when the element \texttt{k} is promoted to level
$h > 0$ because the corresponding \texttt{down} pointer should point to a new
node (due to a split) at level $h-1$ that has not yet been created in a naive
top-down traversal.
%created during a split at the next lower level $h-1$, that has \texttt{k}
%as its header.
Recall from~\secref{prelim} that in a \bskiplist, every header element in a node
at level $i$ must have been promoted to level $i+1$. It is clear which node to
set the \texttt{down} pointer to with a ``bottom-up'' algorithm, because the
corresponding node has been created at level $h-1$ before the \texttt{down}
pointer at level $h$ needs to be set. However, with the proposed ``top-down''
algorithm, the node with the new element \texttt{k} has not yet been created at
level $h-1$ since we are going from $h$ down to $0$.

We can resolve the issue by again exploiting the property of \bskiplist
insertions that the height of every key is determined upfront to \emph{allocate all
  the nodes in advance} that will be created during an insertion. 
% That is, since
% by construction, an element promoted to height $h > 0$  will generate new nodes
% % at levels $0, 1, \ldots, h-1$
That is, when an element is promoted to height $h$, we can allocate $h$ new nodes with \texttt{k} as
the header at the start of any insertion before any interaction with the
skiplist at all. If $h > 1$, we can link these preallocated nodes together in a
stack via \texttt{down} pointers in non-leaf nodes. Therefore, we can fill in
the appropriate slot in the \texttt{down} pointer array at level $h$ with a
pointer to the top of the preallocated stack of new nodes.

To determine where to splice in the other new nodes in levels $0$ to $h - 1$, we continue the traversal level-by-level. Let us consider
the levels in turn, starting with level $h-1$. Just as in the higher
levels, the traversal will follow the \texttt{down} pointer corresponding to the
\texttt{prev} element in level $h$ which will point to some node at level
$h - 1$. At level $h-1$, the traversal will proceed left-to-right until we find
some node \texttt{x} such that \texttt{x->header < k} but
\texttt{x->next->header > k}. Let \texttt{n} be the preallocated node destined
to be at level $h-1$. Elements greater than
\texttt{k} in node \texttt{x} should be moved to \texttt{n}. It is then
linked into the B-skiplist by setting \texttt{n->next = x->next} and
\texttt{x->next = n}. The insertion then proceeds in this way until it reaches
the leaf level by following the \texttt{down} pointer of the last element in
\texttt{x}, which is the largest element at that level less than \texttt{k}.
The element \texttt{k} will appear in the correct nodes with the correct
structure at the end of this top-down method with splits on the way down.

~\figref{bskip-insert} contains a worked out example of a top-down insertion. Notice
how in the example, since the element is promoted to height $h=1$, the
corresponding node at level $0$ is allocated upfront and linked in with
\texttt{down} and \texttt{next} pointers at levels 1 and 0, respectively.

\para{Correctness and bounds}
The remainder of the section will describe the differences resulting from the
top-down insertion algorithm and how they do not affect the \bskiplist's
correctness or asymptotic time bounds.

The main change with the proposed top-down insertion algorithm is a slight
relaxation of the inclusion invariant, or the property that every element present
in level $h$ must also be present at all lower levels, \emph{during}
insertions. It still preserves the property upon every insertion's completion by
linking in the preallocated new nodes via \texttt{down} and \texttt{next}
pointers.

The inclusion invariant is necessary for the \bskiplist to achieve its
asymptotic bounds, because the bounds come from the expected number of levels
and the expected number of nodes traversed in each level. These properties are
derived from the probabilistic coin flips that determine each element's height
in the \bskiplist. However, after each insertion, the pointer structure in the
\bskiplist with the top-down insert algorithm is identical to what it would have
been with a bottom-up algorithm that links in nodes starting from the leaf
level. Therefore, both the number of levels and the number of nodes traversed
per level are unaffected.

\para{Fixed-size nodes} As mentioned in~\secref{intro}, due to the \bskiplist's
randomized structure, any practical \bskiplist implementation must handle the
case where the number of elements in a node exceeds some fixed-size array
allocation.  The number of keys that are supposed to be in a node may exceed a
fixed bound, depending on the sequence of coin flips. Given a \bskiplist with
$n$ elements and promotion probability $p = 1/B$, the worst-case number of
elements in a given node is $\Theta(B\log n)$ w.h.p. Therefore, it is highly
likely that there will be nodes that exceed any fixed size $\Theta(B)$.

The naive solution of table doubling in nodes enables all keys meant for a node
to fit in the corresponding array, but leads to suboptimal performance of
insertions. As mentioned in~\secreftwo{intro}{prelim}, w.h.p., there is a node
in the \bskiplist with $k = \Theta(B\log n)$ elements, so the worst-case
insertion cost in a \bskiplist is $\Theta(\log n)$ cache-line accesses w.h.p.

To alleviate this issue, we modify the \bskiplist design to require fixed-size
node allocations, which may potentially result in node \defn{overflow splits}
(i.e., node splits due to overflow) in addition to \defn{promotion splits} due
to randomization.  If nodes were allowed to grow arbitrarily large, the
worst-case number of elements moved (i.e., shifted to maintain sorted order)
during an insert is $\Theta(B\log n)$, which would take $\Theta(\log n)$ cache
misses. However, by requiring that the nodes have at most $B$ elements, the
number of element moves in any node is at most $B$.

The nodes created from overflow splits do not affect the correctness of inserts
or searches, as operations in the \bskiplist still follow a left-to-right and
top-to-bottom traversal order.  As mentioned in~\secref{intro}, the choice of
fixed-size nodes does not affect the theoretical bounds, because a query would
still have to perform $\Theta(\log n)$ cache-line transfers in the worst case
w.h.p. However, minimizing the cost of inserts is important for practical
efficiency, as element moves must be linear in the node size, while queries can
skip over parts of the node e.g., via binary search.

%%% Local Variables:
%%% mode: latex
%%% TeX-master: "main"
%%% End:

%% file: bskip-concurrency.tex
\section{Top-down concurrency control}\label{sec:concurrency}

This section presents a single-pass \schemename concurrency control scheme for B-skiplists
based on reader-writer locks and the top-down insert scheme described
in~\secref{bskip-insert}. The goal of the proposed scheme is \emph{simplicity} in
%the locking protocol in 
both the \emph{number of top-down traversals} and the
\emph{number of locks held at a time}. Specifically, we will show that each
operation only needs to make a single root-to-leaf traversal. Furthermore, this
traversal only holds a constant number of locks in at most two levels
of the \bskiplist at a time. 
% Due to space reasons, we will include the
% proof of deadlock-freedom in the full version.

% The focus of this section will be on inserts, but deletions are symmetric.

\para{Concurrent finds and range queries}
Let us start with how to implement concurrent finds and range queries with RW
concurrency as an intermediate step to understanding concurrent inserts in
B-skiplists. Since finds and range queries are read-only operations, they only
need to acquire locks in reader mode\footnote{Reader locks are necessary in
  mixed insert-query workloads.}.
% \brian{if we were only worried about these two they don't need locks, its only
% in the presence of inserts that we need locks, not sure if we need to deal
% with this point}
Queries begin on the highest level at the left sentinel and proceed in a
hand-over-hand fashion left-to-right, as described in the HOH scheme for linked
lists in~\secref{prelim}. When the search reaches the node with the appropriate
\texttt{prev} element, it acquires the child node at the next lowest level via
the \texttt{down} pointer using HOH locking in a top-down fashion.  Searches
proceed left-to-right within a level and top-to-down to move between one level
at a time until the query reaches the appropriate node in the leaf level that
should contain the target key. For point finds, the search is then complete and
can release all locks. In contrast, range queries acquire locks left-to-right at
the leaf level in a HOH fashion until the range is exhausted or the search
reaches the end of the skiplist.
%\todo{maybe add a fig}

To recap, both concurrent finds and range queries acquire RW locks in read mode
in a left-to-right and top-to-bottom order in HOH fashion. That is, a thread
holds at most two locks at a time.

\begin{figure}[t]
  \centering \includegraphics[width=\linewidth]{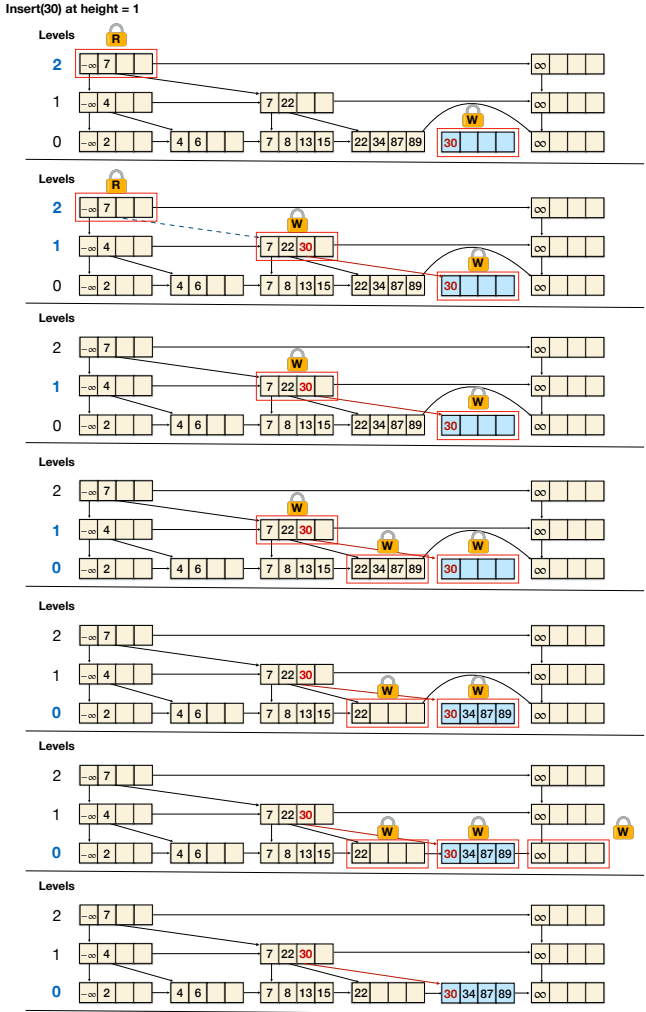}
  \vspace{-.75cm}
  \caption{Example of the insertion of element with key \texttt{k} = 30 and
    height $h = 1$ into a B-skiplist with node size $B = 4$ using the top-down
    CC scheme. The locks with R and W denote acquiring the lock
    in read and write mode, respectively.}
  \label{fig:bskip-concurrent-insert}
\end{figure}

\para{Concurrent inserts}
Next, we will introduce the proposed \schemename concurrency control scheme for
inserts. At a high level, insertions follow a similar left-to-right and
top-to-bottom traversal order to queries. However, inserts raise additional
challenges, since they potentially require structural modification operations
(i.e., splits) if elements are promoted to higher
levels. ~\figref{bskip-concurrent-insert} illustrates a worked example of the
sequence of reader-writer lock acquisitions and node updates with the
\schemename concurrency control protocol.

Just as in the insertion algorithm from~\secref{bskip-insert}, the proposed
\schemename concurrency control scheme leverages the randomized property of
skiplists to complete insertions in \emph{one pass} through the data structure
and to acquire writer locks \emph{only at the levels where writes will
  occur}. That is, it relies on the observation that the level to which an
element is promoted to in a \bskiplist depends only on a sequence of random coin
flips and importantly, is can be determined upfront %at the start of an insert
independently
of the current structure of the skiplist.

Suppose that we are inserting some key \texttt{k} that will be
promoted to level $h$. %as determined by the corresponding sequence of coin
% flips.
Furthermore, suppose that we have preallocated any new nodes that will be
spliced into the skiplist as described in~\secref{bskip-insert}.
% For simplicity, assume that we can take the lock in write mode on all of these
% pre-made nodes.
Since they are not currently linked in the skiplist, taking their write locks
will not delay any other threads.

Inserts begin at the highest level %in the skiplist
and proceed just like queries
for all levels greater than $h$.  Since there will be no writes until level
$h$, the traversal only needs to acquire locks in read mode in the
left-to-right and top-to-bottom traversal until it reaches level $h$.

Once it has reached the appropriate node with the correct \texttt{down} pointer
in level $h + 1$, the insert then acquires the lock on the child node in write
mode on level $h$.  At level $h$, the traversal will continue left-to-right in a
HOH fashion taking locks in write mode until we reach the node where \texttt{k}
should be inserted (i.e., some node \texttt{x} such that \texttt{x->header < k}
and \texttt{x->next->header > k}). Since we have the write lock on \texttt{x},
we can directly insert \texttt{k} in the appropriate slot. If $h = 0$, the
insert is completed, and we can release all of the locks.

If there need to be splits at lower levels (i.e., $h > 0$), we use HOH locking
from top-to-bottom and left-to-right. %to follow the top-down algorithm.
Consider the case of a split at level $h-1$. To find the starting point of the 
traversal at level $h-1$, we first take the write lock on the node pointed to
the by \texttt{down} pointer associated with the \texttt{prev} element in level
$h$. Just as in the single-threaded case, the traversal will then proceed
left-to-right until we find the two nodes that the new preallocated node will
need to be inserted between. However, in the concurrent case, we take writer
locks in HOH fashion left-to-right for thread safety. The actual split mechanism
is unchanged from the serial case, and this process is repeated until the leaf
level.

\para{Integration with fixed-size nodes}
So far, we have described the CC mechanism on logical nodes, but overflow splits do not
affect the guarantees of HOH locking. An insert still only needs to acquire a
constant number of locks at any time. The CC protocol in the
case of overflow splits is even simpler than the case of promotion
splits. Recall that if an element is promoted, the \schemename scheme acquires
at most three locks at once to perform splits: one to update the \texttt{down}
pointer at some level $\ell$ and two to splice in the new node between two
existing nodes at level $\ell - 1$). In overflow splits, there is no need for
the lock to be held on level $\ell$ because there is no \texttt{down} pointer to
the new node, so we only need to hold the two locks on the nodes that we are
splicing the new node between.

\para{Correctness and deadlock-freedom}
The correctness of the proposed HOH-based locking scheme follows directly from
existing theory about HOH locking in linked lists and
\skiplists~\cite{herlihy2020art}. To insert an element into a node without a
split, just acquiring the write lock on that node is sufficient. To split a node
and insert a new node between two existing ones, acquiring the write lock on
both the predecessor node and its \texttt{next} node is sufficient.  There is a
slight relaxation of the maximum number of locks held at once from two to three
during splits. However, locks are held on at most two levels at once.

Finally, the proposed \schemename concurrency scheme is deadlock-free because
there is a \emph{total ordering} on locks from left-to-right within levels and
then top-to-bottom between the levels, thereby avoiding circular wait.

\iffalse
\begin{proof}
The most popular way to avoid deadlock is to ensure that locks are acquired
in a total order, thereby avoiding circular wait. In the proposed top-down
concurrency control protocol, there is a total ordering on the locks from
left-to-right within a level, and then from top-to-bottom in the levels. That
is, the lowest ordered node is the left $-\infty$ sentinel at the highest level,
and the highest ordered node is the right $\infty$ sentinel at the leaf
level. The left $-\infty$ sentinel at some level $\ell$ is ordered directly
after the right $\infty$ sentinel at level $\ell + 1$. The insert and find
algorithms acquire locks according to this total ordering in a left-to-right and
top-to-bottom fashion, thereby avoiding deadlock.
\end{proof}

\fi

%%% Local Variables:
%%% mode: latex
%%% TeX-master: "main"
%%% End:

%% file: eval.tex
\section{Evaluation}\label{sec:eval}
%\eddy{beware all throughput units are in op/us not in op/s, and all latencies are in ns}
% \todo{eval redux with updated numbers on 128 threads + scaling plots}
% \input{latexfigs/micros-table}

This section evaluates the proposed concurrent \bskiplist on the YCSB~\cite{ycsb} compared to several state-of-the-art
concurrent \skiplist- and tree-based indices. As mentioned in~\secref{intro}, we
evaluate all indices in terms of both \emph{throughput} and \emph{latency}. 
% Due
% to space limitations, we omit the raw data tables, but will include them in the
% full version of the paper.

\para{Result summary} At a high level, the \bskiplist achieves
\bskipoverslthroughputrange higher throughput and \bskipoversllatencyrange lower
latency than non-cache-optimized concurrent \skiplists. Furthermore, the
\bskiplist achieves \bskipovertreethroughputpoint throughput on point workloads
and \bskipovertreethroughputrange throughput on range workloads compared to tree-based indices.

\para{Systems setup} All experiments were run on a server with 64-core 2-way
hyperthreaded Intel Xeon Gold 6338 CPU @ 2.00GHz with 1008 GB of memory. The
server has a 3 MiB L1 data cache, a 2 MiB L1 instruction cache, a 80 MiB L2
cache and a 96 MiB L3 cache. We ran all experiments with \numcores physical
cores and \numthreads hyperthreads.

All times are the median of 5 trials after one warm-up trial. To measure
latency, each thread measures the average time taken for a batch of ten
operations\footnote{We measure the average of 10 operations instead of
  individual operation to preserve the contention between threads.} and stores
it in a thread-safe vector. This allows us to sort and calculate the latency at
each percentile after running each benchmark.

We measured the cache misses in~\tabref{intro-cache} with \texttt{perf}.

\para{Workloads} ~\tabref{ycsb-workloads} presents details of the core workloads
from the YCSB~\cite{ycsb-core-workloads}. We tested workloads\footnote{We omit
  workload D from YCSB because it benchmarks the read-latest operation, which is
  not the focus of this work.} A, B, C, and E from the core YCSB workloads
generated with RECIPE~\cite{LeeEtAl19}.

We follow the standard YCSB procedure which consists of two phases: 1) the
\defn{load} phase, where elements are inserted into the data structure, and 2) the
\defn{run} phase, where operations are executed according to the workload's
find/insert ratio. Each workload consists of 100 million (100M) elements
inserted during the load phase, followed by 100M operations executed during the
run phase. All operations in each phase are performed concurrently, but the phases are performed one after another.  %both phases are run concurrently.

We evaluate each workload under both uniform random and zipfian distributions in
the run phase. In the uniform workload, the elements in both the load and run
phases are generated from a uniform distribution.  In the zipfian workload, the
elements in the load phase are generated from a uniform distribution, while the
elements in the run phase are generated from a zipfian distribution. We omit the results on the zipfian distirbution due to space limitations, but the results were similar (on average within $20 \%$) to the uniform distribution. 

%with the
%default YCSB zipfian constant (i.e., theta ~\cite{gray1994quickly} of $0.99$).

% We will include results from both distributions in the full paper.

\input{latexfigs/ycsb_workload_table}

\para{\bskiplist setup} We implemented the concurrent \bskiplist in \texttt{C++}
with an open-source reader-writer lock library~\cite{paralleltools}. The test
driver executes concurrent operations using
pthreads~\cite{nichols1996pthreads}. We ran the \bskiplist with 8-byte keys and
8-byte values (for 16-byte key-value pairs). We set the max height of the
\bskiplist to 5 in our tests.

We compiled the \bskiplist using g++ 11.4.0 with -O3.

\para{\bskiplist sensitivity analysis} We performed a parameter sweep over
promotion probability and node size in the \bskiplist to empirically determine
which settings yield the best performance on the YCSB workloads. 
% Due to space
% limitations, we will include the complete data in the full version of the
% paper. 
Golovin's theoretical paper on \bskiplists proposes a scaling factor,
some constant $c$, on the promotion probability for $p = 1/cB$. Therefore, for
each tested node size we also experiment with $c \in \{0.5, 1.0, 2.0\}$. In
theory, any constant $c$ should suffice to achieve the theoretical randomized
\bskiplist bounds~\cite{golovin2010b}.

Based on the results of this sensitivity analysis, we set the node size in the
\bskiplist to \bskipnodesize bytes (i.e., \bskipnodesizeelts key-value pairs)
and $c = \scalingfactor$, for promotion probability
$p = 1/(\scalingfactor \times \bskipnodesizeelts) = 1/\promotionprob$.

\subsection{Comparison to \skiplist-based indices}

%YCSB + varying insertion percentage, zipf and unif
In this section, we evaluate the \bskiplist against No Hot Spot Skip List
~\cite{crain2013no}, Java \cslm ~\cite{javaskiplist}, and Facebook \folly's
ConcurrentSkipList ~\cite{folly} on the YCSB workloads
\cite{ycsb-core-workloads} and report the results
in~\figreftwo{skiplist-throughput}{skiplist-latency}.

\para{Systems setup} Java \cslm (\javashort) implements a concurrent \skiplist
using a tree-like 2D %two-dimensionally
linked
\skiplist~\cite{javaskiplist-concurrency}.%. It uses a variant of the HM linked ordered set
% algorithm~\cite{javaskiplist-concurrency}.
All operations except range queries are done natively.  We implement range
queries with the \texttt{subMap} interface.

Facebook's \folly (\follyshort) library provides a C++ concurrent skiplist\footnote{%Folly's library:
\url{https://github.com/facebook/folly/blob/main/folly}}. We used the native interface for all point operations and the iterator interface for range queries (since it does not have native support). Folly's skiplist does not support values so we store only the keys.

No Hot Spot Skip List\footnote{
%No Hot Spot Skip List implementation is from
  \url{https://github.com/wangziqi2016/index-microbench/tree/master/nohotspot-skiplist}}
(\nohotshort) is a concurrent, lock free skiplist in C++ that relies on a
background {\it adaptation} thread to maintains the structure of the skiplist
and manage garbage collection. This includes rebalancing the upper index level
to ensure traversals can be done in $O(\log n)$ time. \nohotshort takes in a
{\it sleep time} parameter that determines how frequent the background thread
checks the index and modifies it. In the load phase, we set this parameter to a
relatively small value (100 microseconds) to ensure the index is frequently
balanced. After the load phase, we must wait for the
background thread to balance the height of the tree to $\lg n$
%of the total
%number of keys inserted 
to ensure that operations in the run phase achieve the
desired performance. In the run phase, we set the sleep time higher to 1 second
to ensure the background thread does not stall operations. We do not count the
rebalance time between the load and run phases. %into the load phase time.

We compiled \follyshort and \nohotshort using g++ 11.4.0 with -O3 and Java \cslm
using javac 11.0.25.

 \input{latexfigs/ycsb_skiplist_latency}
 \para{Discussion} ~\figreftwo{skiplist-throughput}{skiplist-latency} illustrate
 the throughputs and various percentile latencies of the
 \skiplists. %~\tabref{skiplist-table} contains the raw data.

 As shown in~\figref{skiplist-throughput}, the
 \bskiplist significantly improves upon the throughput of the other \skiplists on all operations.
 %on  uniform key distributions \eddy{isn't figure 1 is uniform only?}. 
 For the load phase (all inserts), \bskiplist
 achieves about \bsloverjavathroughput, \bslovernhsthroughput, and \bsloverflythroughput\xspace
 %$5\times$, $12\times$, and $3\times$ 
higher throughput,
 respectively, compared to the Java \skiplist (\javashort), the no hot spot
 \skiplist (\nohotshort), and the \folly \skiplist (\follyshort). On point
 workloads (Workloads A, B, and C) with finds/inserts, the \bskiplist is about
\bsloverjavatprange faster than \javashort, \bslovernhstprange faster
 than \nohotshort, and \bsloverflytprange faster than \follyshort. For
 range queries (Workload E), the \bskiplist is about $9\times$ faster than
 \javashort and about $6\times$ faster than \nohotshort and \follyshort.  All
 systems achieves better throughputs on the zipfian workloads than on uniform
 workloads, but the their relative performance to each other remains almost the
 same.

Additionally, as shown in~\figref{skiplist-latency}, the \bskiplist achieves lower latency across all workloads in all tested percentiles (50\%, 90\%, 99\%, and 99.9\%).
%on both uniform and zipfian \eddy{both figure have only uniform data} distributions. 
The most competitive non-blocked \skiplist is the one from \folly, which has at least \flyminlatencyratio and often at least \flycommonlatencyratio higher latency compared to the \bskiplist in the different percentiles.

The \bskiplist achieves better throughput and latency compared to unblocked \skiplists because it reduces  cache misses with better spatial locality in the nodes, as shown in~\tabref{intro-cache}. The folly \skiplist, the fastest of the state-of-the-art \skiplists, incurs between 3.2-5.6$\times$ more cache misses compared to the \bskiplist. Furthermore, the \bskiplist maintains the simple structure and CC schemes that make the \skiplist a popular choice for in-memory indexing. %\eddy{Do we want to update cache miss for 128 t?}

\subsection{Comparison to tree-based indices}
\input{latexfigs/ycsb_tree_throughput}

\para{Systems setup}
We compare the \bskiplist with a high-performance concurrent \bplustree\footnote{
%The Btree OCC implementation can be found in
\url{https://github.com/wheatman/BP-Tree/tree/main/tlx-plain/container}}~\cite{XuLiWh23} (a common \btree variant) based on optimistic
concurrency control~\cite{KungRo81}. The default configuration for the \bplustree sets
$\nodesize=\btreenodesize$ bytes. Both
the concurrent \bskiplist and concurrent \bplustree use the same RW lock.
%implementation.

We also compare the \bskiplist with Masstree\footnote{
%The implementation of Masstree is from
\url{https://github.com/kohler/masstree-beta}}~\cite{MaoKoMo12}, a cache friendly B+-tree variant.
%that incorporates the ideas behind tries.
It utilizes an optimistic concurrency scheme as well. There is native support
for all YCSB operations.
%and all runs were done on default settings.

We compiled the \bplustree and \masstree with g++ 11.4.0 and -O3. %optimization flag. %\todo{add Masstree compiler version}.

\para{Discussion}
~\figreftwo{tree-throughput}{tree-latency} compare the \bskiplist to tree-based
indices on throughput and latency, respectively. We tested both uniform and
zipfian datasets but only illustrate the performance on uniform as the results
are similar.
%The raw data for both can be found in~\tabref{tree-table-small}. We omit the data for the \bskiplist since it is already included in~\tabref{skiplist-table}, but include the performance ratios.

%We find that the \btree achieves consistently higher throughputs and lower latencies than \masstree, so we will focus on it in the following discussion.

Overall, we find that the \bskiplist achieves competitive (between \bskipoverbtreerange higher) throughput compared to the \btree and between \bskipovermasstreerange higher throughput than \masstree on point workloads (load and A-C). We expect the \btree and \bskiplist to be similar on read-heavy workloads because the \bskiplist has a similar height and node size compared to the \btree. Furthermore, the \btree only has to read one node per level, while the \bskiplist may have to take horizontal steps (following \texttt{next} pointers) along each level. Concretely, we found that on average, the \bskiplist takes about 1.7 horizontal steps per level in workloads A-C. As a result, the \bskiplist has a slightly lower throughput (within \bskipwithinbtree) compared to the \btree on workload C.

The \bskiplist achieves the most consistent speedups (\bskipoverbtreeinsertsrange higher throughput) over tree-based indices on insert-heavy workloads (load and A). To understand why, we will first briefly review optimistic concurrency control (OCC)~\cite{KungRo81}, the CC scheme in both the \btree and \masstree. OCC is a classical CC scheme for \btrees based on RW locks that leverages the observation that most insertions only affect the leaf level. Almost all insertions under OCC make one root-to-leaf pass with reader locks on the internal nodes. However, if an element must be promoted, the insert \emph{retires} back to the root, taking write locks on all nodes on the way down. In contrast, the top-down CC scheme in the \bskiplist is guaranteed to always make one pass from root-to-leaf and to take write locks only on the levels that the element is promoted to. To measure this difference, we counted the number of times the root lock was taken in write mode (blocking all other operations) during the load phase and workload A in the \bplustree and \bskiplist. In the load phase, the \bplustree root write lock was taken 26K times, compared to 7 times in the \bskiplist. In workload A, the \btree took the write lock on the root about 8.3K times, compared to 3 times in the \bskiplist. The \bplustree also exhibits higher latency in the 99th percentile compared to the \bskiplist due to the \bplustree's retires back to the root.
\input{latexfigs/latency-micros}
\iffalse
However, since elements may be promoted once they reaelement promotions are only determined once an insert reaches the leaf level in \btrees, an element may have to come ``back up'' once it reaches the bottom. Therefore, OCC makes a first pass from root-to-leaf, taking only read locks on the internal nodes and a write lock only at the leaf. If the leaf is full and an element needs to be promoted, the CC scheme makes a second pessimistic descent, taking write locks from root-to-leaf.
\fi

On the other hand, the \btree achieves about \btreeoverbskipwklde higher throughput than the \bskiplist on range queries (Workload E). As mentioned earlier, although the \bskiplist has $\Theta(B)$ elements per node in expectation, the number of elements actually in a logical node can vary by up to a factor of $O(\log n)$. Therefore, the average density (number of elements in a node) is lower in the \bskiplist compared to the \bplustree, which deterministically splits nodes when they become full. To measure this difference, we counted the number of nodes at the bottom level traversed in both the \bskiplist and \bplustree during workload E. On average, the \bskiplist accesses about $2$ nodes per range query while the \bplustree accesses only about $1.5$ nodes per range query. We note that in-memory indexes are traditionally optimized for point  (OLTP) workloads and range queries are often an optional function. Future work involves improving range queries in \bskiplists by improving the average node density.

\iffalse
\todo{fillin}
On the uniform dataset, \bskiplist is competitive with \bskiplist on throughputs, with \bskiplist winning on load and workload a throughput. In latencies \bskiplist wins in the 50th and 90th percentile latencies on workload e with mostly range queries, with it being 0.6x on both. The other workloads in that range have \bskiplist at 0.9x the latency of B+-Tree. Zipfian shows a similar pattern with the latency differences there being mostly the same.

Compared to Masstree \bskiplist has better load throughputs at around 0.5x. It has much higher throughputs on workload E with range queries, which Masstree is known to suffer on. A, B, and C show moderate throughput increases. \bskiplist also has lower latencies, ranging from 1.1x to 61.6x less on uniform and zipfian. On workload E, \bskiplist is 8.6-11.2x better for both uniform and zipfian. The uniform the load latencies range from being 8.3-17.7x better. Comparatively with zipfian, the 50th and 90th percentiles are around half that of Masstree while the 99th and 99.9th are much lower. On workloads B and C overall, \bskiplist has 1.1-1.7x lower latencies.
\fi

\subsection{Strong Scaling}

% \para{System Setup} This section presents the results of the scaling experiments. We ran \bskiplist, \folly, \btree, \javashort, and \masstree \eddy{do we want to explain why we did not run nohotspot here?} with 1 -- 128 threads. Runs with less than 128 threads were run on a single socket to avoid non-uniform memory access (NUMA) issues across sockets. All systems use the same parameters as before.

%\para{Results}
\figreftwo{scaling_a}{scaling_c} shows the scaling performance of \bskiplist, \folly, \btree, \javashort, \nohotshort, and \masstree on YCSB workload A and C with uniform random keys. \nohotshort scaling starts from 2 threads since it requires at least two threads to run.

On the write-heavy workload A, all systems except \masstree scale number of threads increases. \masstree's performance peaks at 32 threads and declines afterward. On 128 threads, \nohotshort achieves about 26$\times$ speedup. The \bskiplist, \btree, and \folly skiplist all achieve about 35-38$\times$ speedup, while \javashort achieves about $45\times$ speedup. Although \javashort achieves higher parallel scalability, its overall throughput is lower than the C++-based data structures because it does not employ blocking. In contrast, on the read-heavy workload C, all systems achieve higher speedup compared to workload A because there are no writes and therefore less lock contention. On workload C, all systems achieve between 50-60$\times$ speedup on 128 threads except for \nohotshort which achieves about 35$\times$ speedup.

\input{latexfigs/scaling_a}
\input{latexfigs/scaling_c}

%%% Local Variables:
%%% mode: latex
%%% TeX-master: "main"
%%% End:

%% file: latexfigs/ycsb_workload_table.tex
\begin{table}[t]
\centering
\begin{tabular}{@{}ll@{}}
 \hline
Workload & Description \\
 \hline
Load & 100\% inserts from empty\\
A&50\% finds, 50\% inserts \\
B&95\% finds, 5\% inserts \\
C&100\% finds  \\
E&95\% short range iterations (\maxlen = 100), 5\% inserts\ \\
% X&99.9\% finds, 0.1\% inserts \\
%Y&100\% inserts \\ [1ex]
 \hline
\end{tabular}
\caption{YCSB workload descriptions.}
\label{tab:ycsb-workloads}
\vspace{-.5cm}
\end{table}
%%% Local Variables:
%%% mode: latex
%%% TeX-master: "../main"
%%% End:

%% file: latexfigs/ycsb_skiplist_latency.tex
\begin{figure}
  \centering
  \begin{tikzpicture}
    \begin{axis}[
        width=\columnwidth, height=3.5cm,
        axis lines = left,
        xlabel = Percentile,
        ylabel style={align=center, text width=3cm},
        ylabel = {Latency ($\mu$s)},
        %xmode=log,
        ymode=log,
        ymin=1,
      % ymin=1E8,
        ymax=100,
        cycle list name=exotic,
        legend style={at={(0.32, 1.2)}, anchor=north},
        legend columns=2,
        xtick distance=1,
        xticklabels={0, 50, 90, 99, 99.9},xmax=4
      ]

      \addplot
       coordinates {
            (1, 1.39)
            (2, 1.64)
            (3, 2.17)
            (4, 3.22)
        };
      \addlegendentry{B-skiplist}

          \addplot
       coordinates {
            (1, 7.37)
            (2, 9.083)
            (3, 10.820)
            (4, 12.690)
        };
      \addlegendentry{Folly SL}

              \addplot [mark=triangle]
       coordinates {
            (1, 10.200)
            (2, 13.400)
            (3, 15.900)
            (4, 35.000)
        };
      \addlegendentry{Java SL}
              \addplot
           coordinates {
            (1,3.504)
            (2,4.297)
            (3,11.447)
            (4,91.835)
        };
      \addlegendentry{NHS}
    \end{axis}
  \end{tikzpicture}
  \vspace{-.75cm}
  \caption{Latencies of \skiplist-based indices at different percentiles in YCSB workload A with uniform random keys.}
  \label{fig:skiplist-latency}
\end{figure}
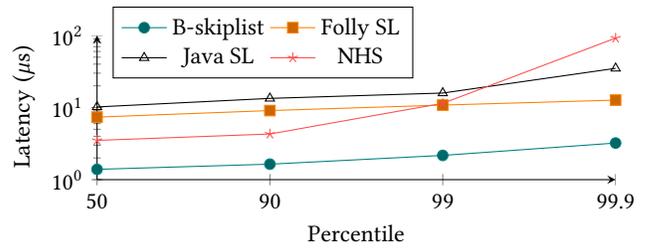
%%% Local Variables:
%%% mode: latex
%%% TeX-master: "../main"
%%% End:

%% file: latexfigs/ycsb_tree_throughput.tex
%\addtocounter{figure}{-2}
    \begin{figure}
  \centering
  \begin{tikzpicture}
    \begin{axis}[
    width=8cm, height=3.5cm,
        ybar,
        legend pos=north west,
        ylabel={Normalized performance},
         ylabel style={align=center, text width=3cm},
        xlabel={Workload},
        xtick=data,
        bar width=6pt,
        legend columns=5,
        extra y ticks=1,
        enlarge x limits=0.14,
        symbolic x coords={Load, A, B, C, E},
        x tick label style={text width = 1cm, align = center},
        ymin = 0,
        ymax = 1.8,
    %    extra y tick style={
    %                 % in case you should remove the grid from the "normal" ticks ...
    %                 ymajorgrids=true,
    %                 % ... but don't show an extra tick (line)
    %                 ytick style={/pgfplots/major tick length=0pt,},
    %                 grid style={violet ,dashed,},
    %         },
    % every axis plot/.append style={fill},
      ]

            \addplot coordinates {
    (Load,1)(A, 1)(B, 1)(C, 1)(E, 1)
    };
      \addlegendentry{\bskiplist}

        \addplot [fill=safe-peach, postaction={pattern=horizontal lines}] table [
          x=workload, y=btree_ratio, col sep=tab
          ] {tsvs/ycsb_uniform_tp.tsv};
          \addlegendentry{B-tree}
        \addplot [] table [
          x=workload, y=masstree_ratio, col sep=tab
          ] {tsvs/ycsb_uniform_tp.tsv};
          \addlegendentry{MassTree}
    \end{axis}
  \end{tikzpicture}
  \vspace{-.25cm}
  \caption{Normalized throughput (ops/s) of tree-based indices relative to the
    \bskiplist with uniform random keys.} %A number below 1 means that the index achieved lower throughput than the \bskiplist.
    \label{fig:tree-throughput}
  \end{figure}
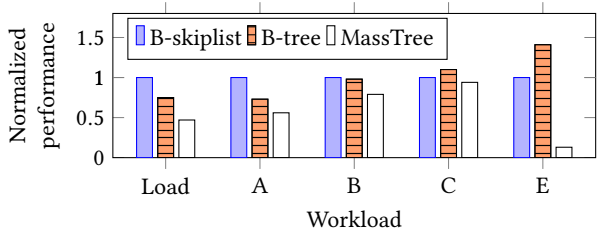

%%% Local Variables:
%%% mode: latex
%%% TeX-master: "../main"
%%% End:

%% file: latexfigs/latency-micros.tex
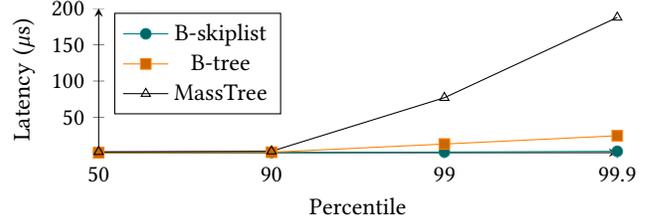
\begin{figure}
  \centering
  \begin{tikzpicture}
    \begin{axis}[
        width=\columnwidth, height=3.5cm,
        axis lines = left,
        xlabel = Percentile,
        ylabel style={align=center, text width=3cm},
        ylabel = {Latency ($\mu$s)},
        % ymode=log,
      % ymin=1E8,
        ymax=200,
        cycle list name=exotic,
        legend pos=north west,
        legend columns=1,
        xtick distance=1,
        xticklabels={0, 50, 90, 99, 99.9},
      ]

      \addplot
       coordinates {
            (1, 1.392)
            (2, 1.641)
            (3, 2.171)
            (4, 3.221)
          %  (5, 3724)
        };
      \addlegendentry{B-skiplist}
              \addplot
               coordinates {
            (1,1.371)
            (2,1.977)
            (3,13.268)
            (4,24.665)
%(5,18775)
        };
      \addlegendentry{B-tree}
              \addplot [mark=triangle]
               coordinates {
(1,2.566)
(2,3.304)
(3,77.013)
(4,187.617)
%(5,170270)
        };
      \addlegendentry{MassTree}

    % \addplot[mark=star, safe-peach,mark options={scale=1.6}]
    %        table[x=insert percentage,y=Bskip 99,col
    %   sep=tab]{tsvs/latencies.tsv};
    %   \addlegendentry{Bskip 99\%}

    %   \addplot
    %  table[x=insert percentage,y=Btree 90,col
    %   sep=tab]{tsvs/latencies.tsv};
    %   \addlegendentry{Btree 90\%}

    %   \addplot[safe-pink, mark=square*]
    %   table[x=insert percentage,y=Btree 99,col
    %   sep=tab]{tsvs/latencies.tsv};
    %   \addlegendentry{Btree 99\%}

    \end{axis}
  \end{tikzpicture}
  \vspace{-.5cm}
  \caption{Percentile latencies of the \bskiplist and tree-based indices on YCSB
    workload A with uniform random keys.}
  \label{fig:tree-latency}
  %\vspace{-.25cm}
\end{figure}
%%% Local Variables:
%%% mode: latex
%%% TeX-master: "../main"
%%% End:

%% file: latexfigs/scaling_a.tex
\begin{figure}[t]
  \centering
  \begin{tikzpicture}
    \begin{axis}[
        width=8cm, height=4.5cm,
        axis lines = left,
        xlabel = Num. Threads,
        ylabel style={align=center},
        ylabel = {Speedup},
        cycle list name=exotic,
        legend pos=north west,
        xmode=log,
        log basis x={2},
        legend columns=2,
        ymax=54,
        ytick={10,20,30,40,50},
        log ticks with fixed point,
        xlabel shift={-2pt},
        xtick={1,2,4,8,16,32,64,128},
        xticklabels = {1,2,4,8,16,32,64,128},
      ]
      \addplot[mark=square, color=red]
      coordinates{
        (1,1.00)
        (2,1.94)
        (4,3.69)
        (8,7.09)
        (16,13.50)
        (32,22.97)
        (64,32.32)
        (128,34.72)
      };
      \addlegendentry{\bskiplist}

    \addplot[mark=triangle, color=blue]
      coordinates{
        (1,1.00)
        (2,2.02)
        (4,3.89)
        (8,7.70)
        (16,14.69)
        (32,26.29)
        (64,36.17)
        (128,37.95)
      };
      \addlegendentry{\folly}   
      \addplot[mark=o, color = green]
      coordinates{
        (1,1)
        (2,1.88)
        (4,3.65)
        (8,6.94)
        (16,13.43)
        (32,23.18)
        (64,30.92)
        (128,45.15)
      };
      \addlegendentry{\javashort}
      
      \addplot[mark=x, color=orange]
      coordinates{
        (1,1.00)
        (2,1.92)
        (4,3.58)
        (8,6.70)
        (16,13.24)
        (32,21.08)
        (64,29.65)
        (128,35.59)
      };
      \addlegendentry{\btree}

      \addplot[mark=+]
      coordinates{
        (1,1.00)
        (2,1.99)
        (4,3.87)
        (8,7.54)
        (16,14.25)
        (32,23.63)
        (64,24.29)
        (128,19.07)
      };
      \addlegendentry{\masstree}

        \addplot[mark=pentagon, color=purple]
      coordinates{
        (2, 1)
        (4,	2.891023897)
        (8,	6.067780216)
        (16,11.50874154)
        (32,16.86111161)
        (64,24.99038491)
        (128,26.01409468)
      };
      \addlegendentry{NoHotSpot}

    \end{axis}
  \end{tikzpicture}
  \vspace{-.5cm}
  \caption{%Run phase throughput s
  Strong scaling 
  of various systems in terms of throughput on YCSB workload A.}% with uniform random keys.}
  \label{fig:scaling_a}
%  \vspace{-.25cm}
\end{figure}
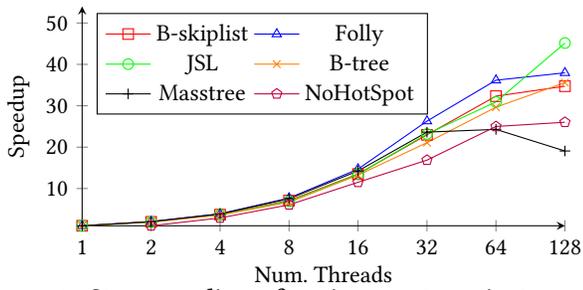

%% file: latexfigs/scaling_c.tex
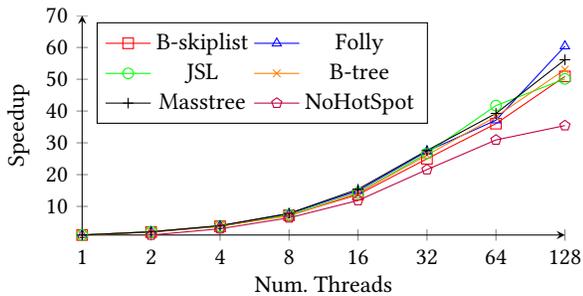
\begin{figure}
  \centering
  \begin{tikzpicture}
    \begin{axis}[
        width=8cm, height=4.5cm,
        axis lines = left,
        xlabel = Num. Threads,
        ylabel style={align=center},
        ylabel = {Speedup},
        cycle list name=exotic,
        legend pos=north west,
        xmode=log,
        log basis x={2},
        legend columns=2,
        ymax=70,
        ytick={10,20,30,40,50,60,70},
        log ticks with fixed point,
        xlabel shift={-2pt},
        xtick={1,2,4,8,16,32,64,128},
        xticklabels = {1,2,4,8,16,32,64,128},
      ]
      \addplot[mark=square, color=red]
      coordinates{
(1,1.00)
(2,1.93)
(4,3.74)
(8,7.22)
(16,13.73)
(32,24.99)
(64,36.03)
(128,50.89)
      };
      \addlegendentry{\bskiplist}

      \addplot[mark=triangle, color=blue]
      coordinates{
(1,1.00)
(2,2.01)
(4,3.90)
(8,7.71)
(16,14.91)
(32,27.34)
(64,37.12)
(128,60.41)
      };
      \addlegendentry{\folly}

      \addplot[mark=o, color = green]
      coordinates{
(1,1)
(2,2.04)
(4,3.71)
(8,7.32)
(16,14.27)
(32,26.72)
(64,41.71)
(128,50.11)
      };
      \addlegendentry{\javashort}
      \addplot[mark=x, color=orange]
      coordinates{
(1,1.00)
(2,1.94)
(4,3.85)
(8,7.01)
(16,14.15)
(32,26.14)
(64,38.24)
(128,53.05)
      };
      \addlegendentry{\btree}
      \addplot[mark=+]
      coordinates{
(1,1)
(2,2.02)
(4,3.95)
(8,7.80)
(16,15.40)
(32,27.59)
(64,39.25)
(128,56.13)
      };
      \addlegendentry{\masstree}
      
      \addplot[mark=pentagon, color=purple]
      coordinates{
(2	, 1)
(4	,2.970190565)
(8	,6.407781824)
(16	,11.86715)
(32	,21.57370374)
(64	,30.90423986)
(128,35.42216312)
      };
      \addlegendentry{NoHotSpot}
      
    \end{axis}
  \end{tikzpicture}
  \vspace{-.5cm}
  \caption{Strong scaling of various systems in terms of throughput on YCSB workload C.}

  \label{fig:scaling_c}
%  \vspace{-.25cm}
\end{figure}

%% file: conclusion.tex
\section{Conclusion}
We present the \bskiplist, a high-performance concurrent in-memory index based
on the \skiplist data structure. The proposed concurrent \bskiplist adapts the
theoretical description of a \bskiplist ~\cite{golovin2010b} for practical
considerations to minimize data movement and mitigate the probabilistic worst
case of element moves. To take advantage of parallel resources, we propose a
top-down insertion algorithm that completes insertion in one pass and a
corresponding simple yet effective CC scheme. The \bskiplist inherits the simple
structure that makes the \skiplist a popular choice for in-memory indexing.

The empirical evaluation demonstrates that the \bskiplist achieves between
\bskipoverslthroughputrange higher throughput and between
\bskipoversllatencyrange lower 99th percentile latency compared to popular
state-of-the-art concurrent \skiplist implementations such as those from
Facebook's folly library and the Java concurrent \skiplist library.
% Furthermore, the \bskiplist achieves similar (\bskipovertreethroughputrange)
% throughput on point workloads and between \bskipovertreelatencyrange lower
% 99th percentile latency on point workloads with inserts compared to
% state-of-the-art tree-based indices.
These results suggest that the \bskiplist is a good candidate for high-performance in-memory indexing because it resolves locality issues in \skiplists while minimizing CC overhead.

For future work, we plan to integrate the \bskiplist into key-value stores like
RocksDB and LevelDB to evaluate its impact on application performance.
Additionally, its high cache locality makes it well-suited for disk-based
indexes.
% , andwe aim to design and implement one using the \bskiplist.
With its strong theoretical guarantees and practical efficiency, we anticipate
that \bskiplist could match or even surpass %the impact of
traditional \skiplists in modern databases.

%Our implementation in C++ is publicaly available at \todo{github link}.

%%% Local Variables:
%%% mode: latex
%%% TeX-master: "main"
%%% End:

%% file: latexfigs/insert_algo.tex
% Insertion algorithm
\begin{algorithm}
\footnotesize
    \caption{Insertion Algorithm}
    \label{alg:insert}
    \begin{algorithmic}[1]
        \Require SkipList $SL$ and a pair key $K$ value $V$
        \Ensure $K$,$V$ inserted into $SL$
        
        \State $h \gets$ level to promote
        \State current $\gets$ starting sentinel
        \State current.\Call{lock}{h, MAXHEIGHT}
        \For{level $\gets$ MAXHEIGHT to 0}
            \State previous$ \gets $current
            \While{current.\Call{next\_header}{} $<=$ key}
                \State current.\Call{next}{}.\Call{lock}{h, level}
                \State previous $\gets$ current
                \State current $\gets$ current.\Call{next}{}
                \State previous.\Call{unlock}{}
            \EndWhile
            \State rank, found $\gets$ current.\Call{find\_key}{$K$}
            \If{found}
                \If{map}
                    \State current.\Call{write}{K,V} at rank
                \ElsIf{set}
                    \State current.\Call{unlock}{} and then \Call{Exit}{}
                \EndIf
            \Else
                \If{h $==$ level}
                    \If{current is overflowing}
                        \State new\_node and then new\_node.\Call{lock}{h, level}
                        \State new\_node.\Call{next}{} $\gets$ current.\Call{next}{}
                        \State current.\Call{next}{} $\gets$ new\_node
                        \State half $\gets$ current.num\_elements/2
                        \State current.\Call{split}{new\_node, half}
                        \If{rank $+ 1 <=$ current.\Call{num\_elements}{}}
                            \State new\_node.\Call{unlock}{}
                        \Else
                            \State current.\Call{unlock}{}
                            \State current $\gets$ new\_node
                            \State rank $\gets$ rank $-$ current.\Call{num\_elements}{}
                        \EndIf
                    \EndIf
                    \State current.\Call{insert}{K,V} at rank$+$1
                    \If{level $>$ 0}
                        \State current.\Call{insert\_child}{rank$+$1}
                    \EndIf
                \ElsIf{h < level}
                    \State new\_node
                    \State new\_node.\Call{next}{} $\gets$ current.\Call{next}{}
                    \State current.\Call{next}{} $\gets$ new\_node
                    \State new\_node.\Call{insert}{K,V} at 0
                    \State current.\Call{split}{new\_node, rank$+$1}
                    \If{level $>$ 0}
                        \State current.\Call{insert\_child}{rank$+$1}
                    \EndIf
                \EndIf
            \EndIf
            \If{level $>$ 0}
                \State previous $\gets$ current
                \State current.\Call{child}{rank}.\Call{lock}{h, level}
                \State current $\gets$ current.\Call{child}{rank}
                \State previous.\Call{unlock}{}
            \EndIf
        \EndFor
    \end{algorithmic}
\end{algorithm}

%% file: latexfigs/locking_logic.tex
% Insertion algorithm
\begin{algorithm}
\footnotesize
    \caption{Lock Logic}
    \label{alg:lock}
    \begin{algorithmic}[1]
        \Require Promotion level $l$, current traversal level $h$, current node $n$
        \Ensure Either read or write locks the current node
        \If{l < h}
            \State n.\Call{read\_lock}{}
        \Else
            \State n.\Call{write\_lock}{}
        \EndIf
    \end{algorithmic}
\end{algorithm}

%% file: latexfigs/zipfian_tp.tex
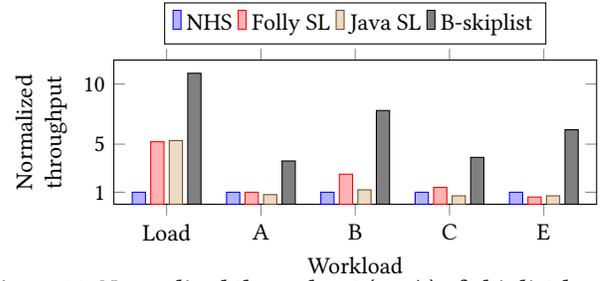
\begin{figure}[H]
  \centering
  \begin{tikzpicture}
    \begin{axis}[
    width=8cm, height=3.5cm,
        ybar,
        legend style={at={(0.5, 1.4)}, anchor=north},
        ylabel={Normalized throughput},
         ylabel style={align=center, text width=3cm},
        xlabel={Workload},
        xtick=data,
        bar width=5pt,
        legend columns=5,
        ytick={1,5,10},
        enlarge x limits=0.14,
        symbolic x coords={Load, A, B, C, E},
        x tick label style={text width = 1cm, align = center},
        ymin = 0,
        ymax = 12,
      ]

    \addplot coordinates {
    (Load,1)(A, 1)(B, 1)(C, 1)(E, 1)
    };
    \addlegendentry{NHS}

     \addplot coordinates {
    (Load,5.2)(A, 1)(B, 2.5)(C, 1.4)(E, 0.6)
    };
    \addlegendentry{Folly SL}

     \addplot coordinates {
    (Load,5.3)(A, 0.8)(B, 1.2)(C, 0.7)(E, 0.7)
    };
    \addlegendentry{Java SL}

     \addplot coordinates {
    (Load,10.9)(A, 3.6)(B, 7.8)(C, 3.9)(E, 6.2)
    };
    \addlegendentry{\bskiplist}

    \end{axis}
  \end{tikzpicture}
  \vspace{-.5cm}
  \caption{Normalized throughput (ops/s) of \skiplist-based indices relative to
    the No Hot Spot Skip List(NHS) ~\cite{crain2013no} under Zipfian key distribution.}
    %A number below 1 means that the index achieved lower throughput than the \bskiplist. \brian{remove second sentance}
    \label{skipziptp}
  \end{figure}

%% file: latexfigs/zipfian_tp_tree.tex
%\addtocounter{figure}{-2}
\vspace{-.7cm}
    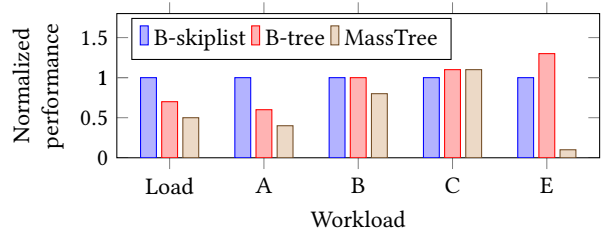
\begin{figure}[H]
  \centering
  \begin{tikzpicture}
    \begin{axis}[
    width=8cm, height=3.5cm,
        ybar,
        legend pos=north west,
        ylabel={Normalized performance},
         ylabel style={align=center, text width=3cm},
        xlabel={Workload},
        xtick=data,
        bar width=6pt,
        legend columns=5,
        extra y ticks=1,
        enlarge x limits=0.14,
        symbolic x coords={Load, A, B, C, E},
        x tick label style={text width = 1cm, align = center},
        ymin = 0,
        ymax = 1.8,
      ]

    \addplot coordinates {
    (Load,1)(A, 1)(B, 1)(C, 1)(E, 1)
    };
    \addlegendentry{\bskiplist}

    \addplot coordinates {
    (Load,0.7)(A, 0.6)(B, 1)(C, 1.1)(E, 1.3)
    };
    \addlegendentry{B-tree}

    \addplot coordinates {
    (Load,0.5)(A, 0.4)(B, 0.8)(C, 1.1)(E, 0.1)
    };
    \addlegendentry{MassTree}

    \end{axis}
  \end{tikzpicture}
  \vspace{-.25cm}
  \caption{Normalized throughput (ops/s) of tree-based indices relative to the
    \bskiplist with zipfian keys.} %A number below 1 means that the index achieved lower throughput than the \bskiplist.
    \label{treeziptp}
  \end{figure}

%% file: latexfigs/zipfian_latency.tex
\begin{figure}[H]
  \centering
  \begin{tikzpicture}
    \begin{axis}[
        width=\columnwidth, height=3.5cm,
        axis lines = left,
        xlabel = Percentile,
        ylabel style={align=center, text width=3cm},
        ylabel = {Latency ($\mu$s)},
        %xmode=log,
        ymode=log,
        ymin=1,
      % ymin=1E8,
        ymax=200,
        cycle list name=exotic,
        legend style={at={(0.32, 1.5)}, anchor=north},
        legend columns=2,
        xtick distance=1,
        xticklabels={0, 50, 90, 99, 99.9},xmax=4
      ]

      \addplot
       coordinates {
            (1, 1.281)
            (2, 1.547)
            (3, 2.054)
            (4, 2.975)
        };
      \addlegendentry{B-skiplist}

          \addplot
       coordinates {
            (1, 8.819)
            (2, 11.229)
            (3, 13.497)
            (4, 15.582)
        };
      \addlegendentry{Folly SL}

              \addplot [mark=triangle]
       coordinates {
            (1, 9.8)
            (2, 12.9)
            (3, 15.3)
            (4, 32.6)
        };
      \addlegendentry{Java SL}
              \addplot
           coordinates {
            (1,3.287)
            (2,3.941)
            (3,8.031)
            (4,44.403)
        };
      \addlegendentry{NHS}

    \addplot
    coordinates {
    (1,1.245)
    (2,1.7)
    (3,12.289)
    (4,25.646)
    };
    \addlegendentry{B-Tree}

    \addplot
    coordinates {
    (1,2.713)
    (2,3.492)
    (3,85.712)
    (4,197.193)
    };
    \addlegendentry{MassTree}
      
    \end{axis}
  \end{tikzpicture}
  \vspace{-.75cm}
  \caption{Latencies of tested indices at different percentiles in YCSB workload A with zipfian keys.}
  \label{ziplatency}
\end{figure}
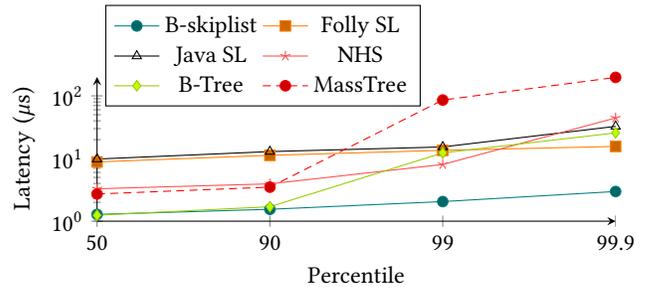

%% file: latexfigs/micros-table.tex
\begin{table*}[t]
  \begin{center}
 %\resizebox{\textwidth}{!}{
%\begin{tabular}{ccccccccccccccccccc}
\begin{tabular}{@{}llllllllllllllllllllll@{}}
  \hline
  & & \multicolumn{8}{c}{\textit{100\% finds}}  & &
                                                    \multicolumn{8}{c}{\textit{100\%
                                                    inserts}} \\
  \cmidrule(lr){4-11}   \cmidrule(lr){13-20}
\textit{Bytes} & \textit {Elts} & \textit{$c$}
%  & \textit{TP} &  \begin{tabular}{@{}c@{}}\textit{Distance} \\ \textit{from
                      %                       best}\end{tabular}
  & \textit{TP} & \textit{DFB}
  & \textit{90\%} &  \textit{DFB}
  & \textit{99\%} &  \textit{DFB}
  & \textit{99.9\%} & \textit{DFB}
  & & \textit{TP} & \textit{DFB}
  & \textit{90\%} &  \textit{DFB}
  & \textit{99\%} &  \textit{DFB}
  & \textit{99.9\%} & \textit{DFB}
  \\
  \hline

&&0.5&28.7&0.6&2.41&0.5&2.8&0.5&4.55&0.4&&13.2&0.4&4.09&0.5&15.45&0.2&284.57&0.0\\
512&32&1&42.5&0.8&1.44&0.8&1.65&0.8&2.16&0.9&&21.2&0.7&3.1&0.7&4.22&0.7&7.09&0.6\\
&&2&37.7&0.7&1.66&0.7&1.94&0.7&2.44&0.8&&14.8&0.5&4.95&0.4&11.23&0.3&20.97&0.2\\
  
  \hline

&&0.5&49.8&1.0&1.23&1.0&1.45&0.9&2&1.0&&29.2&1.0&2.12&1.0&3.09&0.9&4.69&1.0\\
1024&64&1&49.0&1.0&1.28&0.9&1.52&0.9&2.12&0.9&&22.2&0.7&2.62&0.8&5.21&0.6&10.81&0.4\\
&&2&42.1&0.8&1.57&0.7&1.87&0.7&2.55&0.8&&14.9&0.5&4.96&0.4&7.74&0.4&12.58&0.4\\
  
  \hline
  
&&0.5&50.6&1.0&1.17&1.0&1.37&1.0&1.92&1.0&&30.2&1.0&2.02&1.0&2.91&1.0&5.3&0.8\\
2048&128&1&46.6&0.9&1.31&0.9&1.56&0.9&2.2&0.9&&22.9&0.8&2.72&0.7&3.91&0.7&6.39&0.7\\
&&2&43.8&0.9&1.5&0.8&1.76&0.8&2.48&0.8&&14.1&0.5&5.35&0.4&9.88&0.3&15.81&0.3\\

  \hline

&&0.5&45.1&0.9&1.34&0.9&1.57&0.9&2.27&0.8&&25.7&0.9&2.46&0.8&3.32&0.9&4.48&1.0\\
4096&256&1&43.2&0.9&1.44&0.8&1.69&0.8&2.64&0.7&&20.5&0.7&3.25&0.6&5.1&0.6&7.94&0.6\\
&&2&36.3&0.7&1.79&0.7&2.17&0.6&3.48&0.6&&9.6&0.3&10.52&0.2&21.6&0.1&34.06&0.1\\

  \hline

&&0.5&37.7&0.7&1.67&0.7&2.02&0.7&3&0.6&&17.8&0.6&3.81&0.5&5.14&0.6&6.62&0.7\\
8192&512&1&31.8&0.6&2.08&0.6&2.54&0.5&4.32&0.4&&13.8&0.5&5.76&0.4&10.95&0.3&16.81&0.3\\
&&2&33.2&0.7&1.98&0.6&2.35&0.6&3.53&0.5&&10.9&0.4&7.78&0.3&13.86&0.2&19.71&0.2\\

  \hline
\end{tabular}
%}
\end{center}
\caption{
TP = Throughput (in ops/$\mu$s) and $\{90, 99, 99.9\}$ percentile latency (in $\mu$s) of \bskiplist of node sizes from 512 bytes (32 elements per node) to 8912 bytes (512 elements per node), as well as $c = {0.5, 1.0, 2.0}$ in two workloads under a uniform random key distribution. DFB = distance from best (1.0 is the best, all others are normalized).
}
\label{tab:micros}
\vspace{-.5cm}
\end{table*}
%%% Local Variables:
%%% mode: latex
%%% TeX-master: "../main"
%%% End:

%% file: latexfigs/ycsb_skiplist_table.tex
\begin{table*}[t]
  \begin{center}
 \resizebox{\textwidth}{!}{
%\begin{tabular}{ccccccccccccccccccc}
\begin{tabular}{@{}llrrrrrrrrrrrrrrr@{}}
  \hline
  & & \multicolumn{7}{c}{\textit{Uniform}}  & & \multicolumn{7}{c}{\textit{Zipfian}} \\
 \cmidrule(lr){3-9}   \cmidrule(lr){11-17}
   % & & & & \textit{\underline{\javashort}} & & \textit{\underline{\nohotshort}} & & \textit{\underline{\follyshort}} &&&& \textit{\underline{\javashort}} && \textit{\underline{\nohotshort}} && \textit{\underline{\follyshort}}\\\fi
    \textit{Metric} & \textit{YCSB} 
    & \textit{\bskiplistshort}
    & \textit{\javashort}
    & \textit{\javashort/\bskiplistshort} 
    & \textit{\nohotshort} 
    & \textit{\nohotshort/\bskiplistshort} 
    & \textit{\follyshort} 
    & \textit{\follyshort/\bskiplistshort} 
    &&
    \textit{\bskiplistshort}
    & \textit{\javashort}
    & \textit{\javashort/\bskiplistshort} 
    & \textit{\nohotshort} 
    & \textit{\nohotshort/\bskiplistshort} 
    & \textit{\follyshort} 
    & \textit{\follyshort/\bskiplistshort}  \\
      \hline

&Load&20.6&10.0&0.5&1.9&0.1&9.7&0.5&&20.3&9.8&0.5&1.9&0.1&9.7&0.5\\
&A&42.8&9.2&0.2&8.6&0.2&12.8&0.3&&48.7&10.3&0.2&13.5&0.3&13.1&0.3\\
TP&B&75.3&11.5&0.2&8.5&0.1&24.5&0.3&&89.0&13.8&0.2&11.3&0.1&28.7&0.3\\
&C&74.1&11.7&0.2&10.5&0.1&23.0&0.3&&83.3&15.6&0.2&21.3&0.3&29.3&0.4\\
&E&45.3&5.5&0.1&6.1&0.1&5.7&0.1&&54.9&6.2&0.1&8.9&0.2&5.5&0.1\\
        \hline
&Load&1.5&7.2&4.7&52.4&33.8&10.8&7.0&&1.5&7.6&4.9&52.4&33.8&10.8&7.0\\
&A&1.4&10.2&7.3&3.5&2.5&7.4&5.3&&1.3&9.8&7.7&3.3&2.6&8.8&6.9\\
50\%&B&1.2&9.6&8.2&3.4&2.9&4.6&3.9&&1.0&9.5&9.5&2.8&2.8&3.9&3.9\\
&C&1.2&9.8&8.5&3.5&3.0&4.3&3.8&&1.0&8.4&8.7&2.7&2.8&3.7&3.8\\
&E&1.8&13.8&7.6&12.3&6.7&16.7&9.2&&1.6&13.5&8.6&9.5&6.0&20.3&12.9\\
\hline
&Load&1.9&9.8&5.2&112.1&59.1&13.8&7.3&&1.9&10.0&5.3&112.1&59.1&13.8&7.3\\
&A&1.6&13.4&8.2&4.3&2.6&4.3&2.6&&1.5&12.9&8.3&3.9&2.5&3.9&2.5\\
90\%&B&1.3&12.6&9.6&4.5&3.4&5.2&3.9&&1.2&12.0&10.2&3.7&3.2&4.6&3.9\\
&C&1.3&12.6&9.6&4.7&3.6&4.9&3.8&&1.1&11.1&9.8&3.6&3.1&4.3&3.8\\
&E&2.1&20.5&9.9&20.0&9.7&37.2&17.9&&1.8&18.5&10.1&16.2&8.8&27.8&15.1\\
\hline
&Load&2.5&9.8&3.9&257.5&103.2&16.9&6.8&&2.5&10.0&4.0&257.5&103.2&16.9&6.8\\
&A&2.2&15.9&7.3&91.8&42.3&10.8&5.0&&2.1&15.3&7.4&44.4&21.6&13.5&6.6\\
99\%&B&1.7&14.4&8.4&154.2&89.9&6.0&3.5&&1.5&13.8&9.2&77.5&51.6&5.5&3.6\\
&C&1.5&14.5&9.4&150.0&97.5&5.6&3.7&&1.3&13.0&9.7&76.1&57.0&5.1&3.8\\
&E&2.5&29.9&12.1&165.5&67.2&47.2&19.2&&2.2&22.7&10.3&88.5&40.0&33.7&15.3\\
\hline
&Load&51.3&35.0&0.7&982.4&19.1&20.0&0.4&&51.3&32.6&0.6&982.4&19.1&20.0&0.4\\
&A&3.2&26.0&8.1&91.8&28.5&12.7&3.9&&3.0&51.4&17.3&44.4&14.9&15.6&5.2\\
99.9\%&B&2.3&26.0&11.1&154.2&65.9&9.6&4.1&&2.1&51.4&24.1&77.5&36.4&9.2&4.3\\
&C&2.2&56.8&26.0&150.0&68.8&9.2&4.2&&2.0&16.2&8.3&76.1&38.8&8.6&4.4\\
&E&2.2&536.8&246.1&150.0&68.8&55.9&25.6&&2.0&330.0&168.3&76.1&38.8&41.3&21.1\\
\iffalse
\hline
&Load&83.0&25.5&0.3&70.6&0.9&10.7&0.1&&93.3&25.5&0.3&85.7&1.0&9.6&0.1\\
&A&3.7&2945.1&790.8&645.4&173.3&11.3&3.0&&3.1&1961.5&526.7&326.2&87.6&11.8&3.2\\
99.99\%&B&2.6&1608.6&630.6&986.3&386.6&10.7&4.2&&2.4&1607.0&629.9&512.2&200.8&10.1&4.0\\
&C&2.3&1608.1&712.5&966.2&428.1&11.3&5.0&&2.1&1605.8&711.5&943.0&417.8&10.8&4.8\\
&E&3.7&1029.7&274.8&976.7&260.7&25.3&6.7&&3.4&94.3&25.2&501.9&134.0&27.8&7.4\\
\fi
        \hline
        
\end{tabular}
}
\end{center}
\caption{
% Throughput (in operations/s) \todo{and latencies} of the \bskiplist (\bskiplistshort), Java
%   \skiplist (\javashort), optimistic \btree
%   (\btreeshort), and \masstree (\masstreeshort) on
%   uniform random workloads from YCSB. A number above 1 means that the system achieved higher throughput than the \bskiplist. \todo{add another one with zipf}\todo{also,
%   maybe one of each of these for 90\%, 99\%, 99.9\%?}
Throughput (TP, in ops/$\mu$s) and $\{90, 99, 99.9\}$ percentile latency (in $\mu$s) of the following \skiplist-based indices: \bskiplist (\bskiplistshort), Java \skiplist (\javashort), No hot spot \skiplist (\nohotshort), and folly's \skiplist (\follyshort). All measurements are normalized against the \bskiplist. For throughput, a number below 1 in the ratio means that the index achieved a lower throughput compared to the \bskiplist. For latency, a number below 1 in the ratio means that the index achieved a lower latency than the \bskiplist. 
}
\label{tab:skiplist-table}
\vspace{-1cm}
\end{table*}

%%% Local Variables:
%%% mode: latex
%%% TeX-master: "../main"
%%% End:

%% file: latexfigs/ycsb_tree_table.tex
\begin{table}[t]
  \begin{center}
\resizebox{\columnwidth}{!}{
%\begin{tabular}{ccccccccccccccccccc}
\begin{tabular}{@{}llrrrrrrrrrrr@{}}
  \hline
  & & \multicolumn{4}{c}{\textit{Uniform}}  & & \multicolumn{4}{c}{\textit{Zipfian}} \\
 \cmidrule(lr){3-6}   \cmidrule(lr){8-11}
    & & & \textit{\underline{\btreeshort}} & & \textit{\underline{\masstreeshort}} &&& \textit{\underline{\btreeshort}} && \textit{\underline{\masstreeshort}} \\
    \textit{Metric} & \textit{YCSB} 
    %& \textit{\bskiplistshort}
    & \textit{\btreeshort}
    & \textit{\bskiplistshort} 
    & \textit{\masstreeshort} 
    & \textit{\bskiplistshort} 
    &
    %& \textit{\bskiplistshort}
    & \textit{\btreeshort}& \textit{\bskiplistshort} & \textit{\masstreeshort} 
    & \textit{\bskiplistshort} \\
    \hline 
&Load&15.5&0.8&9.6&0.5&&14.5&0.7&9.6&0.5\\		
&A&31.2&0.7&24.0&0.6&&30.5&0.6&21.7&0.4\\		
TP&B&75.3&1.0&60.7&0.8&&90.7&1.0&72.9&0.8\\		
&C&81.6&1.1&69.7&0.9&&88.7&1.1&94.8&1.1\\		
&E&64.0&1.4&6.0&0.1&&68.9&1.3&6.9&0.1\\		
        \hline
&Load&1.9&1.3&3.6&2.4&&19.5&12.6&36.4&23.5\\
&A&1.4&1.0&2.6&1.8&&12.5&9.7&27.1&21.2\\
50\%&B&1.1&0.9&1.5&1.3&&8.7&8.7&12.5&12.5\\
&C&1.0&0.9&1.5&1.3&&8.4&8.7&12.1&12.6\\
&E&1.3&0.7&21.4&11.8&&10.7&6.8&181.7&116.0\\
        \hline
&Load&12.6&6.6&5.6&3.0&&125.6&66.2&56.2&29.6\\
&A&2.0&1.2&3.3&2.0&&17.0&11.0&34.9&22.6\\
90\%&B&1.2&0.9&1.6&1.2&&10.3&8.8&15.0&12.8\\
&C&1.1&0.9&1.6&1.3&&9.8&8.7&14.1&12.4\\
&E&1.4&0.7&26.6&12.8&&12.6&6.9&232.3&126.5\\
        \hline
&Load&33.2&13.3&160.4&64.3&&331.7&132.9&1604.5&643.1\\
&A&13.3&6.1&77.0&35.5&&122.9&59.8&857.1&417.3\\
99\%&B&3.2&1.9&2.4&1.4&&18.2&12.1&22.2&14.8\\
&C&1.3&0.9&2.2&1.5&&11.4&8.5&20.0&15.0\\
&E&2.7&1.1&30.7&12.5&&30.3&13.7&274.5&124.2\\
        \hline
&Load&122.6&2.4&410.4&8.0&&1226.4&23.9&4104.3&79.9\\
&A&24.7&7.7&187.6&58.2&&256.5&86.2&1971.9&662.8\\
99.9\%&B&12.0&5.1&3.0&1.3&&97.8&45.9&24.2&11.3\\
&C&2.0&0.9&2.4&1.1&&17.0&8.6&23.5&12.0\\
&E&12.2&5.6&35.2&16.2&&109.8&56.0&307.1&156.6\\
\hline
\end{tabular}
}
\end{center}
\caption{
Throughput (TP, in ops/$\mu$s) and $\{90, 99, 99.9\}$ percentile latency (in ns) of the following tree-based indices: the optimistic \btree (\btreeshort) and Masstree (\masstreeshort). All measurements are normalized against the \bskiplist (\bskiplistshort). For throughput, a number below 1 in the ratio means that the index achieved a lower throughput compared to the \bskiplist. For latency, a number below 1 in the ratio means that the index achieved a lower latency than the \bskiplist.
}
\label{tab:tree-table-small}
\end{table}

%%% Local Variables:
%%% mode: latex
%%% TeX-master: "../main"
%%% End: